\documentclass[12pt,preprint]{aastex}

\newcommand{\etal}{et al.}
\newcommand{\ie}{i.e.}
\newcommand{\eg}{e.g.}

\def\spose#1{\hbox to 0pt{#1\hss}}    
\def\ltsim{\mathrel{\spose{\lower 3pt\hbox{$\mathchar"218$}}
     \raise 2.0pt\hbox{$\mathchar"13C$}}}
\def\gtsim{\mathrel{\spose{\lower 3pt\hbox{$\mathchar"218$}}
     \raise 2.0pt\hbox{$\mathchar"13E$}}}

\shortauthors{Norton, Gebhardt, Zabludoff, Zaritsky}

\shorttitle{Internal Kinematics of E+A Galaxies}

\begin{document}

\title{The Spatial Distribution and Kinematics of \\ 
Stellar Populations in E+A Galaxies}

\author{Stuart A. Norton\altaffilmark{1}, Karl Gebhardt\altaffilmark{1,2}, Ann I. Zabludoff\altaffilmark{3} and Dennis Zaritsky\altaffilmark{3}}

\altaffiltext{1}{University of California Observatories/Lick
Observatory, Board of Studies in Astronomy and Astrophysics,
University of California, Santa Cruz, CA 95064; stuart@ucolick.org}

\altaffiltext{2}{Astronomy Department, University of Texas at Austin,
Mail Code C1400, Austin, TX  78712; gebhardt@astro.as.utexas.edu}

\altaffiltext{3}{Steward Observatory, University of Arizona, 933 N. Cherry
Street, Tucson, AZ 85721; azabludoff@as.arizona.edu, dzaritsky@as.arizona.edu}

\begin{abstract}

We use long-slit spectroscopic observations of the sample of E+A
galaxies described by Zabludoff~\etal\ to constrain the nature of the
progenitors and remnants of the E+A phase of galaxy evolution.  We
measure spatially-resolved kinematic properties of the young ($\ltsim
1$ Gyr) and old ($\gtsim$ few Gyr) stellar populations.  The young
stellar populations are more centrally concentrated than the older
populations, but they are not confined to the galaxy core (radius
$\ltsim$ 1 kpc).  The kinematics of the old stellar population place
16 of 20 of our E+As on a trend parallel to the Faber-Jackson relation
that is offset by $\sim$ 0.6 mag in R.  Eighteen of 20 E+As have
${\rm v}/\sigma < 1$.  As the young stars in these systems evolve, the
luminosity offset will disappear and the remnants will be
pressure-supported systems that lie on the Faber-Jackson relation.
Although Zabludoff~\etal\ spectroscopically selected the most extreme
E+A galaxies in the local volume, the sample is kinematically diverse:
velocity dispersions range from $\lesssim$ 30 km s$^{-1}$ to $\sim$
200 km s$^{-1}$ over a luminosity range of M$_{\rm R}$ = $-19$ to $-22
+ 5$ log $h$. Combining these results with an estimate of the number
of galaxies that experience an E+A phase, we conclude that the E+A
phase of galaxy evolution is important in the development of a large
fraction of spheroid-dominated galaxies over a wide range of
luminosities and masses.  Our kinematic observations, together with
evidence that E+As have recently evolved from a vigorous star forming
phase to a quiescent phase (\eg,\ Couch and Sharples, Caldwell~\etal)
and that many have tidal features consistent with disk-like
progenitors (Zabludoff~\etal), indicate that these galaxies are
undergoing a transformation from gas-rich, star-forming,
rotationally-supported, disk-dominated galaxies into gas-poor,
quiescent, pressure-supported, spheroid-dominated galaxies.

\end{abstract}



\section{Introduction}

One of the oldest questions in galaxy evolution is the evolutionary
connection, if any, between spiral and elliptical galaxies.  More
specifically, is it possible to identify unambiguously a population of
gas-poor, quiescent, pressure-supported, spheroid-dominated galaxies
that were formerly gas-rich, star forming, rotation-supported, disky
galaxies?  The spectra of ``E+A'' galaxies\footnote{The term ``E+A''
comes from their spectra, which contain absorption lines from an old
stellar population similar to that of elliptical galaxies (E-type) in
addition to a relatively young (A-type) stellar population.  We use
the E+A designation for historical reasons only and note that the
purely spectroscopic designation K+A (\eg,\ Franx 1993) is more
appropriate.} \citep{dg83}, which have a relatively young (A-type)
stellar population but lack significant [OII] emission, indicate that
these galaxies had high levels of star formation $\sim 1 - 2$ Gyr ago
\citep{cs87,nbk90,schweizer96} and that star formation has effectively
stopped within that time.  Thus, E+As may be a crucial link between
gas-rich, star forming galaxies and quiescent galaxies.  Some E+As
have tidal features (Zabludoff~\etal\ 1996; hereafter Z96) roughly
consistent with simulations of the merger of two spiral galaxies
\citep{t2}, suggesting that E+As lie morphologically between the end of
the Toomre merger sequence \citep{toomre77} and a relaxed spheroidal
end-product.

While E+A galaxies possess some spectroscopic and morphological
signatures of an evolutionary transition, their internal kinematics
are poorly understood --- most known E+As lie at intermediate and high
redshifts.  As a result, it has been impossible to answer the most
important questions concerning this seemingly major transition phase
in galaxy evolution: Do the kinematics of E+As support the picture in
which star formation is abruptly extinguished in otherwise normal disk
galaxies or are they consistent with the galaxy merger models that
form spheroids?  If E+As are pressure-supported, then are their
properties consistent with those of normal spheroidal galaxies?  What
is the luminosity range of galaxies that could have evolved through an
E+A phase?  With long-slit spectroscopy of a substantial
sample of nearby E+As, we can address these issues for the first time.

The 21 E+As drawn from 11,113 galaxies in the Las Campanas Redshift
Survey (LCRS; Shectman~\etal\ 1992, 1996) by Z96 is
the largest uniformly-selected sample of E+As within $z \sim 0.2$.
These galaxies have strong Balmer absorption features (average
equivalent width of H$\beta$, H$\gamma$, and H$\delta > 5.5$ \AA) and
little if any [OII] emission (flux detections are at a level $\leq
2\sigma$).  The majority ($\sim 75\%$) of these E+As lie in the field,
well outside rich cluster environments.  In this paper, we determine
how the fraction of A stars, the stellar velocity, and the stellar
velocity dispersion vary along the major axis for these galaxies.

What triggers the starburst and subsequent cessation of star
formation in E+A galaxies are puzzles.  Possible explanations include
galaxy-galaxy mergers or interactions
\citep{schweizer82,lh88,lk95,schweizer96}, interactions of galaxies
with the intracluster medium \citep{dg83,bd86}, and galaxy harassment
\citep{moore98}.  Detections of tidal features (Schweizer 1996, Z96)
and the discovery of E+As outside the cluster environment
(\cite{ohh91}, Z96) suggest that mergers or interactions are indeed at
least one way to produce E+As.  To understand whether this mechanism
is responsible for all E+A's, we must examine their stellar kinematics
and their spatial distribution of young stars.  If major mergers are
the dominant mechanism responsible for E+A formation, and merger
remnants settle quickly to the virial plane (Mihos 1999), we should
find that a large fraction of E+As are dynamically pressure-supported
and lie within the region of parameter space --- the Faber-Jackson
relation \citep{fj76} --- defined by normal spheroidal galaxies.  In
contrast, disk-like kinematics in E+As would support a model in which
star formation is abruptly extinguished in otherwise normal disk
galaxies.  Franx (1993) finds strong rotation in one distant, cluster
E+A, and Caldwell~\etal\ (1996) report rotation in three E+As in the
Coma cluster. Unfortunately, these studies were limited to cluster
environments, the samples are small, and the criteria for selecting
E+As differ among authors (cf. \S~\ref{sec:discussion}). We present
kinematic results for a sample of uniformly and conservatively defined
E+As that span a range of environments.

Complementing the kinematics, the spatial distributions of the young
and old stellar populations further discriminate among formation
models. For example, if E+As are otherwise normal late-type galaxies
in which star formation is quickly extinguished, one expects little or
no difference between the distributions of old and young stars. On the
other hand, if mergers are occurring, simulations predict a high
concentration of young stars toward the center because gas is driven
inward after dissipating its orbital energy
\citep{noguchi88,bh91,mrb92,mh96}.  Star formation in mergers and
their remnants is not limited to the nucleus
\citep{rieke85,wright88,schweizer96,mb98} as suggested by some
numerical simulations, but it is centrally concentrated.  Observations
of several E+A galaxies in clusters suggest that young stars are
present throughout the galaxy \citep{franx93,crfl96}.  It is not
known, however, if these trends are characteristic of all E+As,
including those outside of rich clusters.  With spatially-resolved
spectroscopy spanning a wide range in wavelength, we can now
separate the distributions and kinematics of the young and old stellar
populations in the Las Campanas E+A sample.

If the E+A phase does mark a transition from disk to spheroid, what
range of spheroids can be produced?  The velocity dispersions of
normal ellipticals range from $\sim 50$ to $\sim 300$ km s$^{-1}$
\citep{defis,bn90}, but the distribution for E+As is unknown.  In
lower luminosity ellipticals rotational support is increasingly
important \citep{bn90}, but no test for this trend has been made with
E+A galaxy samples.  If E+As occupy only a limited part of the
Faber-Jackson relation then they mark an interesting, but not
necessarily significant, phase in the evolution of some galaxies.
However, if we observe E+As over the range of luminosities and
velocity dispersions occupied by normal ellipticals, then this brief
phase may be part of the evolution of many galaxies.  Examining the
kinematic properties of this large, unbiased sample of E+As will help
us determine not only whether the E+A phase represents a transition
between galaxy types, but also what fraction of galaxies are likely to
have evolved through it.

A concern in the interpretation of E+A spectra is the degree to which
bursts of star formation may be cloaked by dust or missed by the often
small physical radius probed by the spectroscopic fiber or slit.
\citet{smail99} conclude, based upon observations at near infrared and
radio wavelengths, that some distant ``post-starburst'' galaxies are
actually dust-obscured starbursts.  To determine whether significant star
formation is enshrouded by dust in our E+A sample, two studies (Chang
et al. 2001, Miller et al. in preparation) search for radio continuum
emission, which can be interpreted as evidence for on-going star
formation and is impervious to dust. Of the E+As in the Z96 sample,
only two of the 17 observed to date are detected in radio continuum.
Their detection limits of a few times $10^{21}$ $h^{-2}$ W Hz$^{-1}$
correspond to upper limits of 0.5 -- 1 $h^{-2}$ M$_{\sun}$ yr$^{-1}$.
Thus, we conclude that E+As are not disguised starbursts.


To better understand the properties and origins of E+As along the
lines just described, we obtain and analyze long-slit spectra for the
21 E+A galaxies in the Z96 sample.  In
Section~\ref{sec:data} we describe the data, our reduction procedure,
and our methods for separating the kinematics and spatial
distributions of young and old stellar populations.  In
Section~\ref{sec:results} we present the results of our kinematic
analysis.  In Section~\ref{sec:discussion}, we discuss the
implications of our results, and in Section~\ref{sec:conclusions} we
present our conclusions. Throughout this analysis we assume a flat,
$\Omega_0$ = 1 cosmology and H$_0$ = 100 km s$^{-1}$ Mpc$^{-1}$,
which results in an image scale of $\sim$ 1 kpc per arcsec for the
typical galaxy in our sample.


\section{Data and Analysis}

\label{sec:data}

\subsection{Observations and Reduction}

We obtained long-slit spectra of our sample of E+As using the du Pont
2.5-meter telescope at Las Campanas\footnote{Las Campanas observatory
is operated by the Observatories of the Carnegie Institution of
Washington.} during November 1995, February 1996, and April 1996.  We
used the B\&C Spectrograph with the TEK-1 (1024 pixels $\times$ 1024
pixels) CCD.  Table~\ref{tab:observations} summarizes our
observations: galaxy name and position (Columns 1--3), the total
exposure time (Column 4), and the spectral range (Column 5). Galaxy
names, EA~1 through EA~21, denote increasing strength of the 4000\AA\
break spectral feature (cf. Z96). Galaxies with lowest numbers have
the weakest break and thus have had the most recent and/or intense
burst of star formation.  For the remainder of this paper, we refer to
each galaxy by its assigned number.

\begin{deluxetable}{ccccc}
\tablewidth{413.265pt}
\tablecaption{Observations \label{tab:observations}}
\tablehead{
\colhead{Galaxy\tablenotemark{a}} &
\colhead{RA} &
\colhead{Dec} &
\colhead{Total t$_{exp}$ (s)} &
\colhead{Observed $\lambda$ range (\AA)}
}
\startdata
1  & 10$^{\rm h}$58$^{\rm m}$48\fs 97 & $-$11\degr 54\arcmin 9\farcs 80 &  7200 & 4320 -- 5630 \\                              
2  & 2$^{\rm h}$15$^{\rm m}$43\fs 24  & $-$44\degr 46\arcmin 36\farcs 70 &  8100 & 4450 -- 5760 \\                           
3  & 12$^{\rm h}$6$^{\rm m}$31\fs 34  & $-$12\degr 5\arcmin 55\farcs 40 &   5400 & 4320 -- 5630 \\                           
4  & 3$^{\rm h}$58$^{\rm m}$23\fs 42  & $-$44\degr 43\arcmin 40\farcs 29 &  4400 & 4490 -- 5810 \\                           
5  & 1$^{\rm h}$56$^{\rm m}$0\fs 12   & $-$44\degr 51\arcmin 49\farcs 0 &    9000 & 4540 -- 5850 \\                           
6  & 11$^{\rm h}$51$^{\rm m}$21\fs 96 & $-$2\degr 53\arcmin 55\farcs 9 &   7200 & 4380 -- 5680 \\                           
7  & 22$^{\rm h}$38$^{\rm m}$17\fs 77 & $-$38\degr 50\arcmin 17\farcs 0 &  5400 & 4490 -- 5810 \\                           
8  & 14$^{\rm h}$29$^{\rm m}$20\fs 26 & $-$12\degr 44\arcmin 18\farcs 20 & 7200 & 4500 -- 5810 \\                           
9  & 1$^{\rm h}$15$^{\rm m}$24\fs 20  & $-$41\degr 50\arcmin 10\farcs 69 &  5400 & 4310 -- 5620 \\                           
10 & 2$^{\rm h}$9$^{\rm m}$44\fs 50   & $-$44\degr 21\arcmin 43\farcs 20 &   7200 & 4490 -- 5810 \\                           
11 & 1$^{\rm h}$12$^{\rm m}$34\fs 56  & $-$41\degr 38\arcmin 21\farcs 79 &  7200 & 4540 -- 5850 \\                           
12 & 12$^{\rm h}$3$^{\rm m}$25\fs 97  & $-$2\degr 37\arcmin 50\farcs 59 &   6000 & 4450 -- 5760 \\                           
13 & 11$^{\rm h}$17$^{\rm m}$21\fs 50 & $-$12\degr 36\arcmin 13\farcs 30 & 5400 & 4450 -- 5760 \\                           
14 & 13$^{\rm h}$54$^{\rm m}$20\fs 95 & $-$12\degr 12\arcmin 10\farcs 60 & 3600 & 4380 -- 5680 \\                           
15 & 14$^{\rm h}$38$^{\rm m}$5\fs 59  & $-$6\degr 27\arcmin 4\farcs 90 &    7200 & 4540 -- 5850 \\                           
16 & 12$^{\rm h}$17$^{\rm m}$21\fs 44 & $-$5\degr 57\arcmin 22\farcs 70 &  5400 & 4380 -- 5680 \\                           
17 & 10$^{\rm h}$11$^{\rm m}$20\fs 17 & $-$2\degr 40\arcmin 53\farcs 0 &   7200 & 4320 -- 5630 \\                           
18 & 0$^{\rm h}$20$^{\rm m}$18\fs 81  & $-$41\degr 50\arcmin 15\farcs 60 &  5400 & 4310 -- 5620 \\                           
19 & 2$^{\rm h}$5$^{\rm m}$51\fs 66   & $-$45\degr 35\arcmin 2\farcs 79 &    5700 & 4320 -- 5630 \\                           
20 & 0$^{\rm h}$36$^{\rm m}$20\fs 9   & $-$39\degr 13\arcmin 41\farcs 50 &   3600 & 4310 -- 5620 \\                           
20\tablenotemark{b} & 0$^{\rm h}$36$^{\rm m}$20\fs 9 & $-$39\degr 13\arcmin 41\farcs 50 & 3600 & 4310 -- 5620 \\    
21  & 11$^{\rm h}$12$^{\rm m}$52\fs 65 & $-$6\degr 28\arcmin 51\farcs 59 & 7200 & 4450 -- 5760 \\

\enddata 

\tablenotetext{a}{Numbers identifying E+A galaxies are from
Z96.}  

\tablenotetext{b}{This spectrum was taken with the slit
along the galaxy minor axis.}

\end{deluxetable}

The data for this study consist of three to four coadded 1800 second
exposures of each E+A galaxy using a 1.5-arcsecond slit.  The pixel
scale is 0.59 $\arcsec$/pixel in the spatial direction and 1.3
\AA/pixel in spectral direction. The spatial resolution, measured from
stellar spectra, is $\sim 1.6\arcsec$, and typically varies by $\sim
\pm 0.3\arcsec$ during each night.  The spectral resolution, measured
from the night sky lines, is 1.95 pixels full-width at half maximum
(or 2.5 \AA) and typically varies by $\sim$ 5 \% between observations.
Because one of our principal goals is to examine both the older and
younger stellar populations, the spectral range is selected to include
both Mg I 5175\AA, which mostly originates from the older populations,
and H$\delta$, which mostly originates from the younger populations.
The total spectral coverage is $\sim 1300$ \AA.  The resulting spectra
have a signal-to-noise (S/N) of $\sim$ 40 per pixel for an aperture
along the slit that includes the entire galaxy.  In all cases, we
aligned the slit with the major axis of the galaxy; for EA 20, we also
obtained a minor-axis profile.



We apply standard IRAF\footnote{IRAF, the Image Reduction and
Analysis Facility, is distributed by the National Optical Astronomy
Observatories, which are operated by the Association of Universities
for Research in Astronomy, Inc., under cooperative agreement with the
National Science Foundation.} routines and reduction techniques to the
raw images.  First, we subtract both an
overscan region and an average bias frame from each image.
Next, we divide the image
by an image of a uniformly illuminated region of the
dome interior. We obtain separate flat-field images for
each grating angle that we use.
We remove small-scale variations in
sensitivity, and preserve the large-scale illumination function and
slit function, by fitting and dividing the flats by a 2-dimensional
function (25th order in the wavelength direction and 8th order in the
spatial direction).  We obtain spectra of twilight skies to
correct for the slit illumination function in the object spectra.  We
derive the illumination pattern by fitting the sky spectra with a 7th
order function in the spatial direction at five different points
along the wavelength axis.  We do not correct the CCD response
function in the wavelength direction.  After removing the sensitivity
and illumination variations, a number of pixels remain that 
are consistently discrepant from the surrounding pixels in all images. We
identify these CCD defects interactively and interpolate over each group of
bad pixels in the narrowest direction spanned by the pixels.

Defects are introduced into individual images by cosmic rays.  We
apply an algorithm similar to that described by \citet{windetal94} to
remove these. For each galaxy, the stack of 2D spectra (typically,
three) are compared by examining the pixel values within a moving
boxcar.  We calculate the median in the boxcar and use the CCD noise
characteristics to determine the expected standard deviation within
the boxcar.  We then compare the value of the pixel at the center of
the boxcar among the images in the stack.  All pixels greater than a
threshold deviation (typically 12 to 15 $\sigma$) from the boxcar
median are replaced by the median.  We visually examine the resulting
images and tune the threshold deviation to ensure that no emission
lines are removed.  Finally, we average the image stack to produce the
final image.

After removing cosmic rays, we complete the reduction process by
establishing the wavelength scale, rectifying the images, and
subtracting the sky flux.  The reference spectra for wavelength
calibration are helium-argon lamp images taken before and after each
galaxy image.  We use a two-dimensional fit (7th order spatially and
spectrally) from the calibration spectra to wavelength calibrate and
rectify each galaxy spectrum.  Finally, we fit a line in the spatial
direction to the sky on either side of the slit and subtract the sky
spectrum.  Due to the angular length of the long slit (3.5 arcmin) and
the small angular extent of the galaxies (several arcseconds), the
edges of the frames provide uncontaminated sky spectra.


\subsection{Measuring Line-of-Sight Velocity Distributions and Stellar Populations}
\label{sec:fits}
The observed spectral range, which includes H$\beta$, H$\gamma$,
H$\delta$, Mg I, and the G-band, enables us to extract the velocity
distributions of young and old stars. Although the different spectral
features are {\it primarily} associated with either young or old
populations, they do not arise exclusively in one or the other.  It
would be incorrect, for example, to measure the velocity distribution
of the young stellar population using the Balmer lines, while ignoring
the contribution of the older population to those lines.  To
disentangle the kinematics of the two stellar populations, we {\em
simultaneously} fit the velocity profile and the relative contribution
of different stellar components. Our two major concerns in the
extraction of the line-of-sight velocity distribution (LOSVD) are
including the appropriate stellar templates and optimizing the
combined stellar templates in fitting the observed spectra.  We
discuss these in turn.

The library of template spectra must contain the range of stellar spectral types
found in the galaxy and match the spectral resolution and wavelength range of the
galaxy spectrum.
Ideally, one would construct a large template library with the
same instrumental configuration as used for the galaxy observation. However, because of the
redshift difference between galaxies and template stars, we could not
match the spectral coverage of local standard stars to that of the E+A
galaxies (which lie at $0.06  \ltsim z \ltsim 0.12$) without incurring
problematic overhead.  Instead, we use publicly available
spectra of template stars 
from the Coud\'e Feed Spectral
Library at KPNO \citep{leith96} and convolve those spectra to match 
the characteristics of our
observational configuration.  The KPNO library contains 684 stars with a
broad range of spectral types at 1.8~\AA\ FWHM resolution. We select a
subsample of 22 stars representing A dwarfs and G and K dwarfs and
giants.  We experimented with a variety of different stars, and our
final template subsample provides good overall type coverage without
involving an exorbitant number of stars. Note that A giants are not
included among the templates because their lifetimes ( $<$ 1 Myr) are
extremely short relative to main-sequence A star lifetimes 
($\sim$ 1 Gyr; \citet{bressan93}).

\begin{figure}
\epsscale{0.9}
\plotone{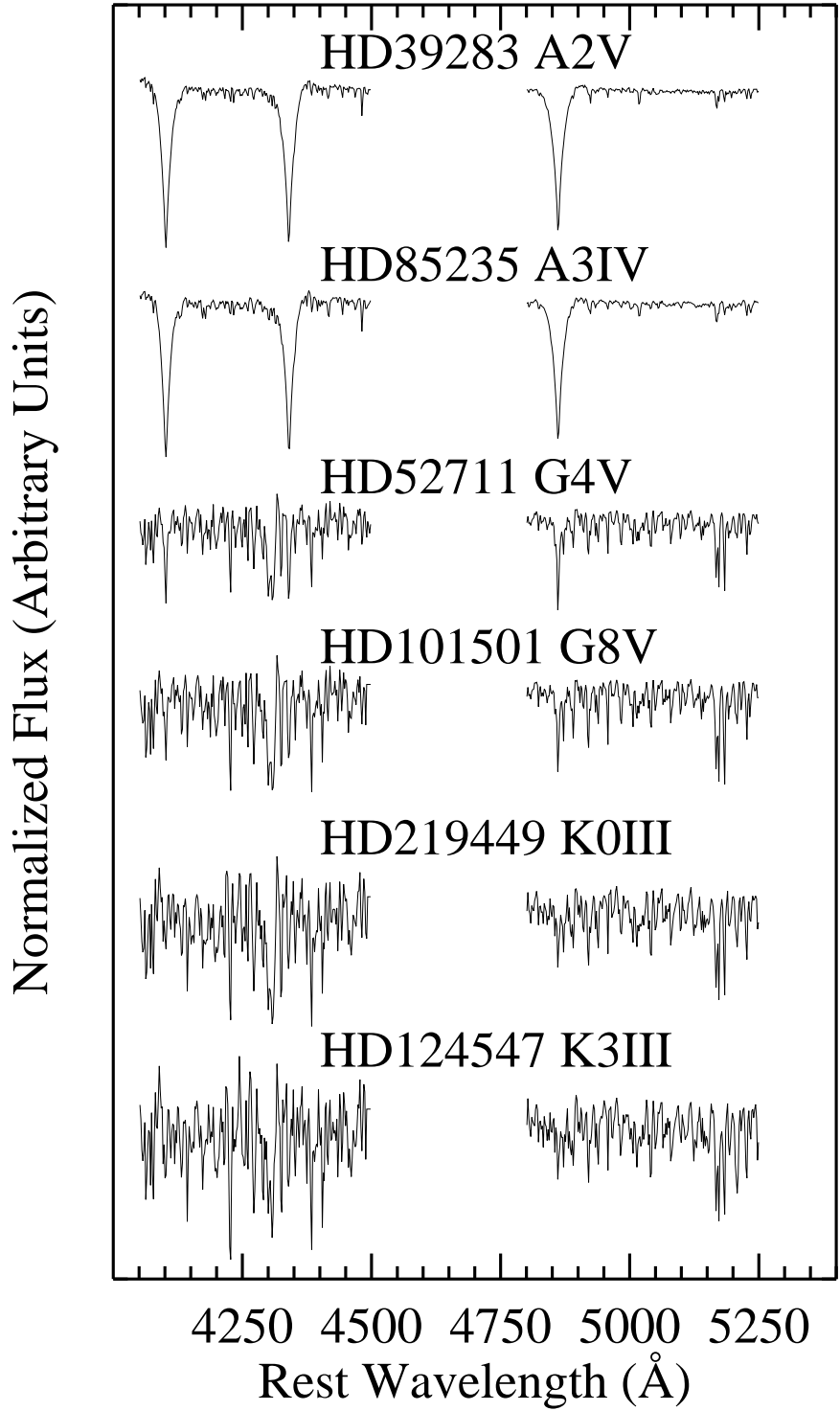}
\caption{Representative template star spectra from
\citet{leith96}.  Each spectrum is labeled with its spectral
type.\label{fig:templatespec}}
\end{figure}

In Figure~\ref{fig:templatespec}, we plot representative template
spectra.  It is apparent from the figure that the most significant
spectral differences are between the A-type stars and the other
templates.  We correspondingly separate the templates into two groups:
a ``young'' population represented by the A dwarfs, and an ``old''
population represented by the G- and K-type templates.  The presence of main
sequence A stars, whose lifetimes are $\sim$ 1 Gyr, and the lack of 
bluer stars in the E+A spectra  indicate
that the most recent star formation event occurred $\sim$ 1 Gyr ago.
The main sequence stars comprising the
older population have lifetimes from 4 to 10 Gyr.  We refer
to the two stellar populations as the young and old populations.

To prepare the template stars for their application,
we transform each spectrum to the resolution and
redshift of each galaxy spectrum.  We convolve the spectra using a
Gaussian (\eg,\ \citet{vdmf93}) with $\sigma^2 =
\sigma^2_{\rm g} - \sigma^2_{\rm t}$.  The
resolution of the galaxy spectrum is $\sigma_{\rm g}$ (1.1~\AA) and
that of the templates is $\sigma_{\rm t}$ (0.9~\AA\ for the KPNO
library).

We measure the LOSVD from a series
of five apertures along the slit and from an aperture that
contains all of the galaxy light. The five apertures include
a central aperture that is 1 pixel wide, two apertures at intermediate
radii that are adjacent to the central aperture and are 2
pixels wide, and two more distant apertures that are
adjacent to the intermediate apertures and are 3 pixels wide (\ie,
the most distant aperture on each side of the slit spans the region
from 2.5 to 5.5 pixels [or $\sim$ 1.5 to 3.3 kpc] from the center).
Next, we fit and divide by the continuum in all template and galaxy
spectra, alleviating the need for accurate flux calibration of our
spectra.

We use a maximum likelihood procedure to estimate the best-fitting
Gaussian LOSVD and template weights.  We begin by choosing an initial
velocity profile for each template type (young or old), which is
 characterized
by the fractional contribution from each template star to the total
light and by the Gaussian parameters (mean velocity and velocity
dispersion) for that population. We convolve this line profile with a
weighted-averaged template (the average of the template spectra where
each star has its particular weight), subtract this spectrum
from the galaxy spectrum, and calculate the residuals. We vary the
template weights and velocity profile parameters (Gaussian
coefficients) and identify the best match to the galaxy spectrum.
Each template type is forced to have uniform kinematics,
\ie, all of the young, A-type stars have the same LOSVD, but the
contribution of each A star template to the total spectrum is allowed
to vary. Rather than allowing the relative contributions of the populations
to vary freely, we could have used evolutionary spectral synthesis codes
(\eg,\ \citet{bc93}) to model the contributions.
However, because the integrated model spectra for populations of age $\ltsim$
1 Gyr are rather uncertain and because such a procedure introduces other
unknowns such as the initial mass function, we prefer to fit the relative
contributions of young and old stars to the spectrum.

We vary portions of this analysis to test our assumptions.  For
example, we add skew and kurtosis terms by replacing the Gaussian
profile LOSVD profile with a Gauss-Hermite profile.  We find no
significant differences between the Gaussian and Gauss-Hermite fits,
and thus we do not report the results for the Gauss-Hermite fits.  We
also compare the results from our standard method to those from a
non-parametric fit \citep{getal01} and find that the two methods yield
similar results. We conclude that deviations from a Gaussian are
sufficiently insignificant that they do not adversely affect the
results of the parametric fits.

We perform a Monte-Carlo analysis to measure the uncertainties in the
recovered parameters.  We generate 100 Monte Carlo realizations of the
data in each aperture by adding random noise with the appropriate RMS
to the best-fit spectrum (which is constructed from the best-fitting
combination of templates and the LOSVD).  Each realization is then
analyzed as described above.  From the distribution of 100 model
parameters, the 16th and 84th percentile values determine the 68\%
confidence band.  The average squared difference between the
best-fitting model and the data provides the RMS. This approach does
not account for systematic differences between the models and
observations, such as template mismatches.  For this same reason, we
are unable to accurately determine uncertainties in the young star
light fraction when the best-fitting value is zero, in which case we
adopt a minimum error of 0.1.

\subsection{Spectral Fitting}

\begin{figure}
\plotone{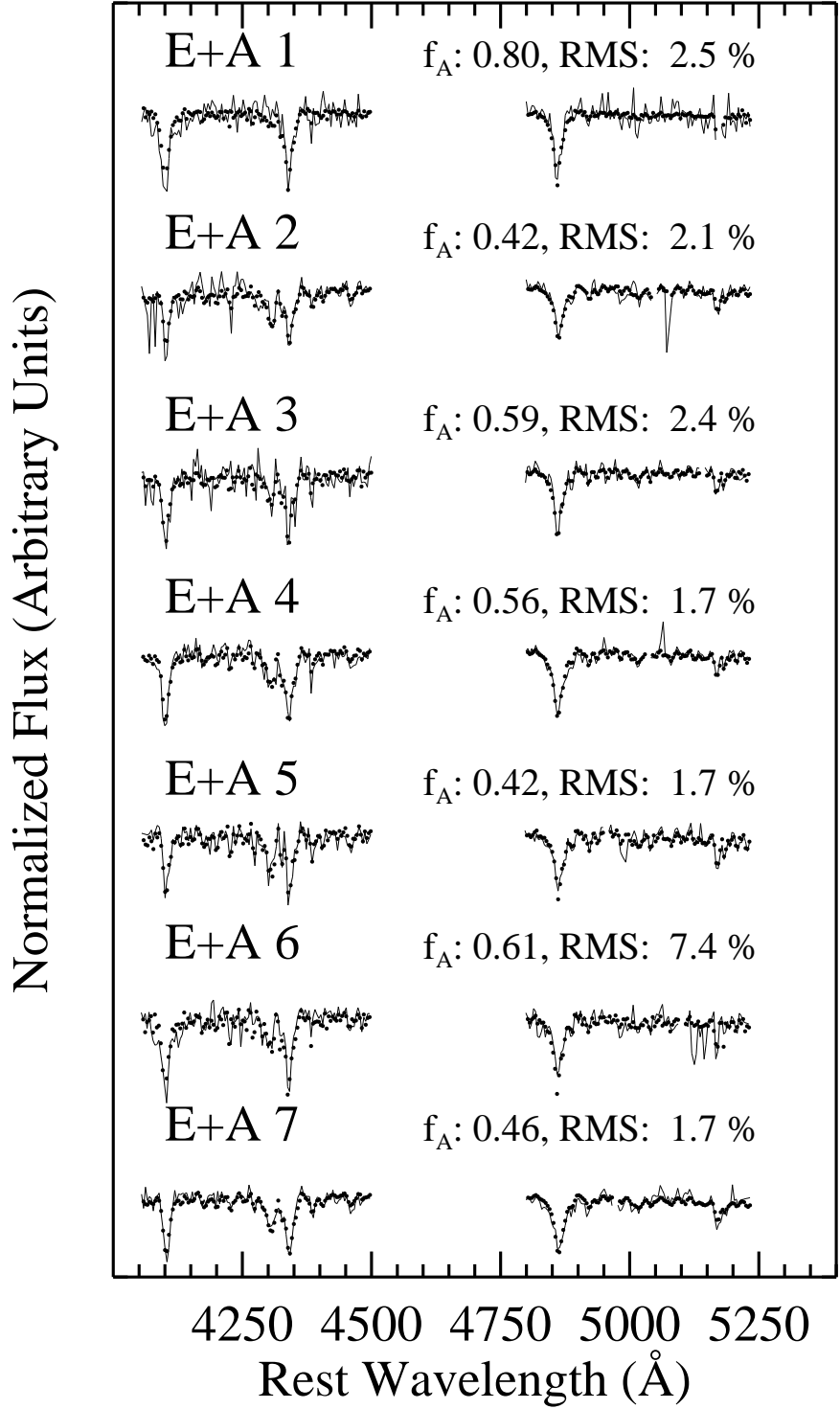}
\caption{Observed spectra (solid lines) and our
best fit (points).  The RMS (per pixel) for each spectrum is
calculated using the deviation between the continuum-normalized fit
and data. \label{fig:firstfits}}
\end{figure}

\begin{figure}
\plotone{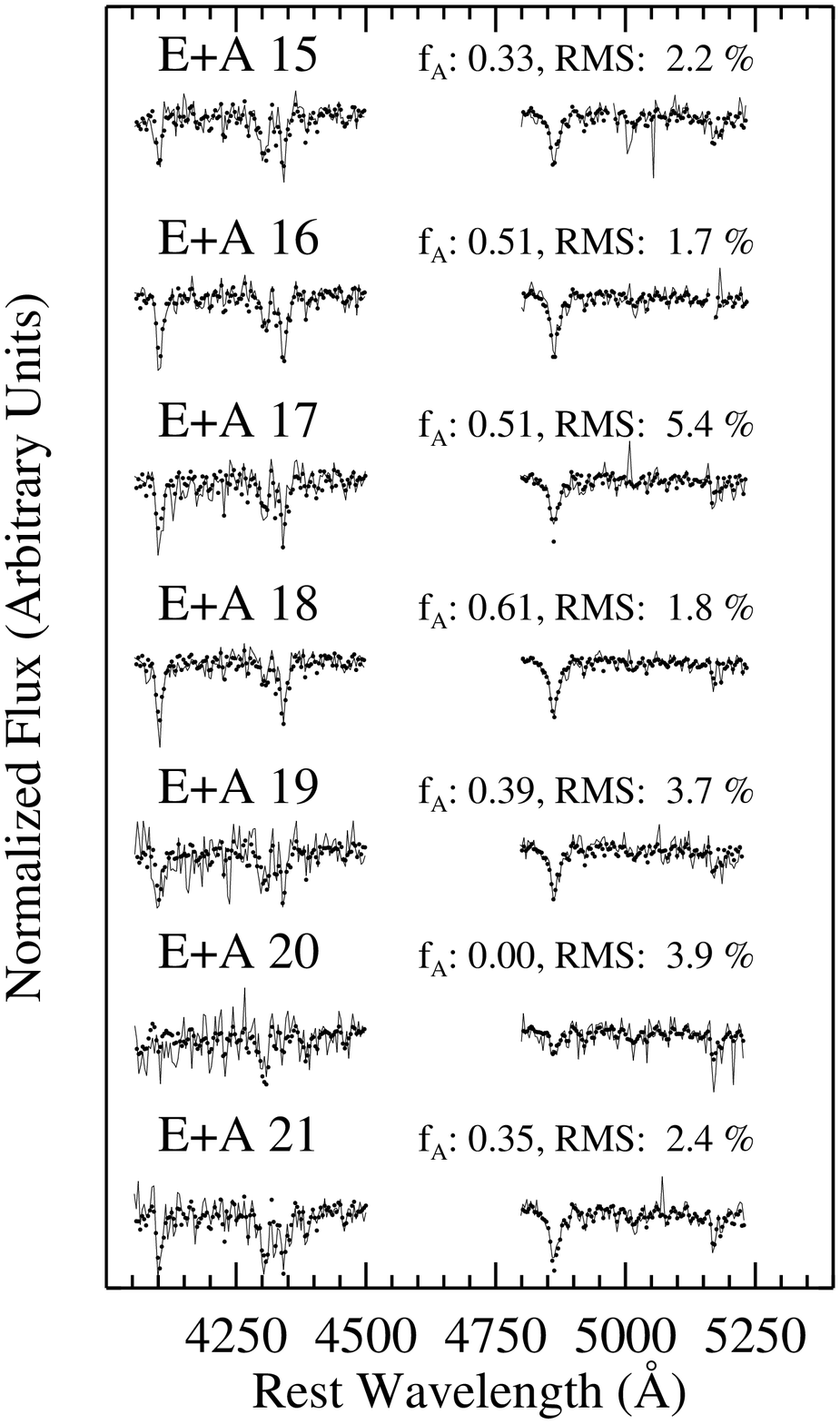}
\caption{Observed spectra (solid lines) and our
best fit (points).  The RMS (per pixel) for each spectrum is
calculated using the deviation between the continuum-normalized fit
and data.\label{fig:lastfits}}
\end{figure}

We present examples of the fitted spectra in
Figures~\ref{fig:firstfits} through~\ref{fig:lastfits}.  The observed
spectrum of each galaxy, for the full ($\sim 6.6 \arcsec$) aperture,
is plotted along with the best-fit model.  The RMS difference between
the fit and the data is labeled, along with the fraction of light from
A-type (young) stars, above each spectrum.

\begin{figure}
\plotone{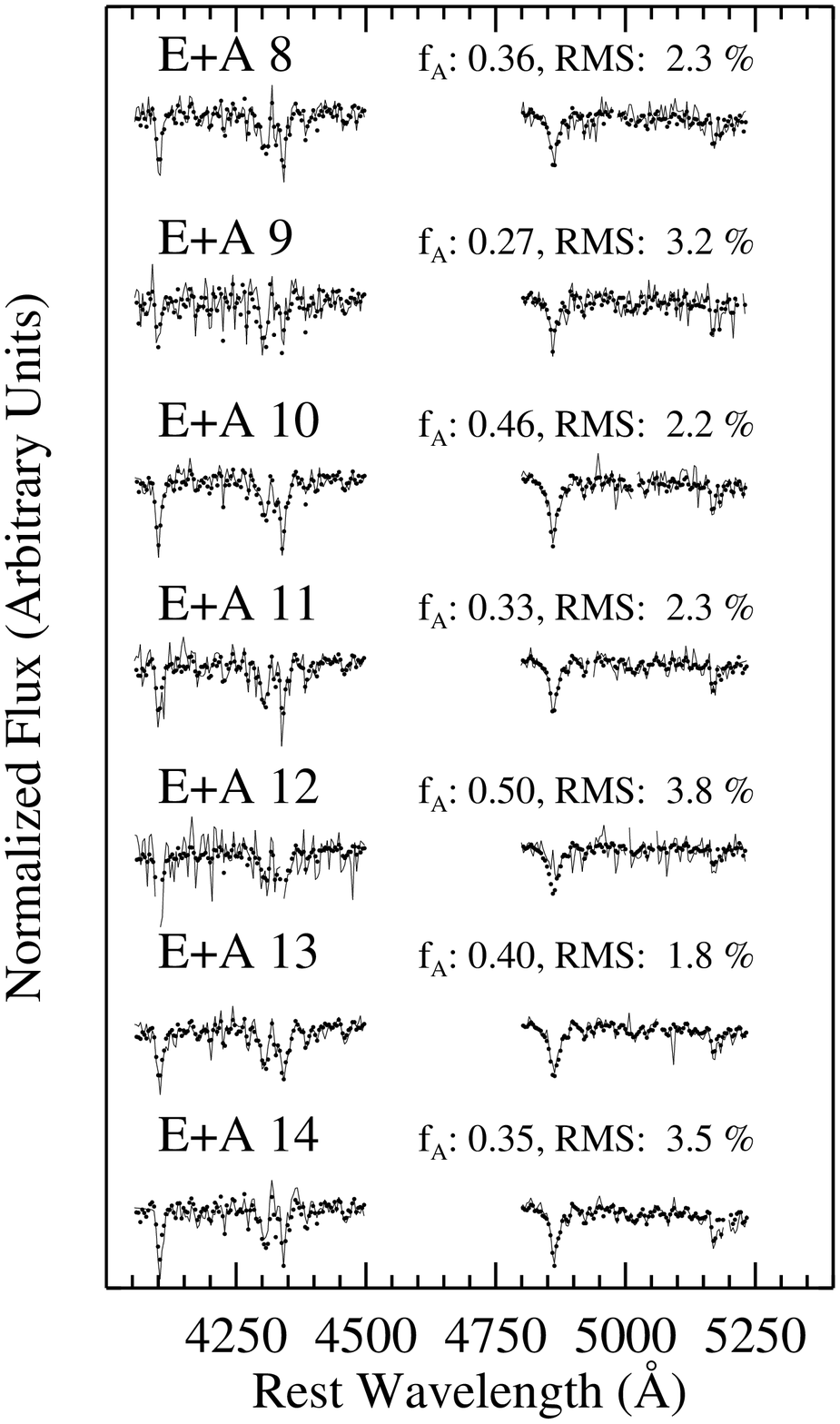}
\caption{Observed spectra (solid lines) and our
best fit (points).  The RMS (per pixel) for each spectrum is
calculated using the deviation between the continuum-normalized fit
and data.}
\end{figure}

An examination of the full set of spectra reveals a minor, but 
systematic trend: in
the majority of cases, the best-fitting model has a deficit of
H$\delta$ absorption, the correct amount of H$\gamma$, and an excess
of H$\beta$ absorption.  The most likely cause
of this discrepancy is the lack of an accurate flux calibration. 
Because we have divided the spectra of both galaxies
and templates by their continua, and because the relative contributions
of continuum and lines vary differentially for young and old templates 
across our wavelength range, we are diluting the absorption features 
differently in the templates than in the galaxies. For example,
because the younger
stars have more blue continuum
than the older stars, we are artificially diluting
the bluer Balmer lines and enhancing the redder lines when we divide the
young star spectra by their continua. We could
rectify this situation by having our algorithm fit the continuum 
as well as the line strengths (which would
require flux calibration of both template and galaxy spectra),
but given the small differences seen in the equivalent widths of
the Balmer lines (cf. Figures~\ref{fig:firstfits}--\ref{fig:lastfits}) we
conclude that the potential biases from the uncertainties in 
the flux calibration are at least as large as those from the 
Balmer discrepancies. Other potential sources for the
discrepancy include an incomplete template library and/or emission-line
filling of the Balmer absorption lines. We have checked both of these
possibilities and  neither provides a satisfactory explanation. 
These small discrepancies between the fitted and observed spectra
have an insignificant effect on the
kinematics, which is the primary focus here.

\begin{figure}
\epsscale{0.8}
\plotone{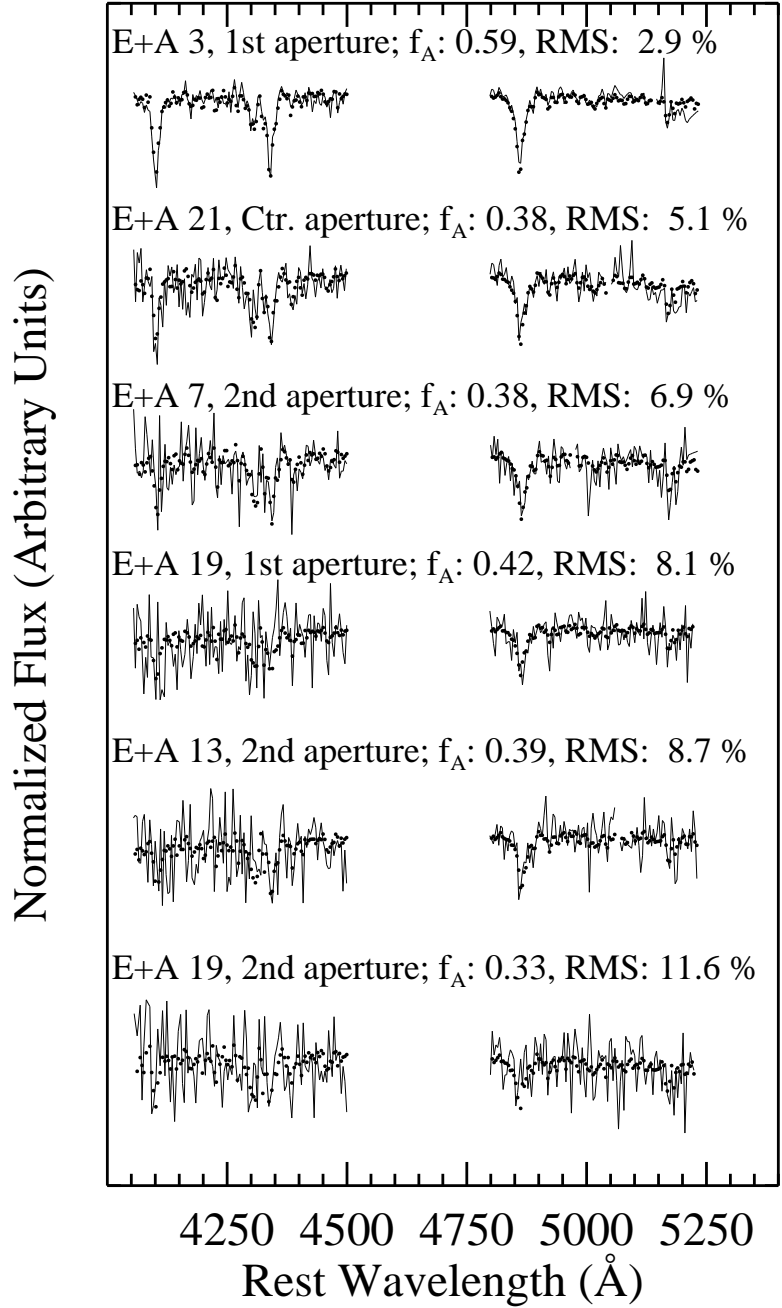}
\caption{Representative spectral fits at various radii.  From upper to
lower panels these plots show: EA~3 at a radial distance of
0.9$\arcsec$ ($\sim$ 0.9 kpc), the central aperture for EA~21, EA~7 at
a radial distance of 2.4$\arcsec$ (3.4 kpc), EA~9 at 0.9$\arcsec$ (0.8
kpc), EA~13 at 2.4$\arcsec$ (2.9 kpc), and EA~19 at 2.4$\arcsec$ (2.0
kpc).  Note that the last spectrum is below our signal-to-noise cutoff
(RMS $>$ 10\%) and is not used in the analysis. The RMS is calculated
as in Figure~\ref{fig:firstfits}. \label{fig:fitexamples}}
\epsscale{0.9}
\end{figure}

As mentioned previously, to investigate the properties of each
galaxy as a function of radius, we have also extracted spectra within
several radial bins.  In Figure~\ref{fig:fitexamples}, we show six
examples of the fits in these radial bins, spanning a representative
S/N range, for the central aperture and the two radial
apertures.  The spectrum in the final panel, which comes from the 
outermost aperture in EA~19, is obviously too
poor for a kinematic analysis.
We make no attempt to derive kinematics for this
spectrum or for any aperture with a lower S/N; this limit corresponds
to S/N $\sim$ 10 or RMS $\sim$ 0.1 (10\% of the continuum flux).
However, our Monte-Carlo simulations demonstrate that the 
young star fraction is a significantly more robust measurement than
the kinematics, and we retain these measurements for all five
apertures of each galaxy.

\section{Results}

\label{sec:results}

\begin{figure}
\plotone{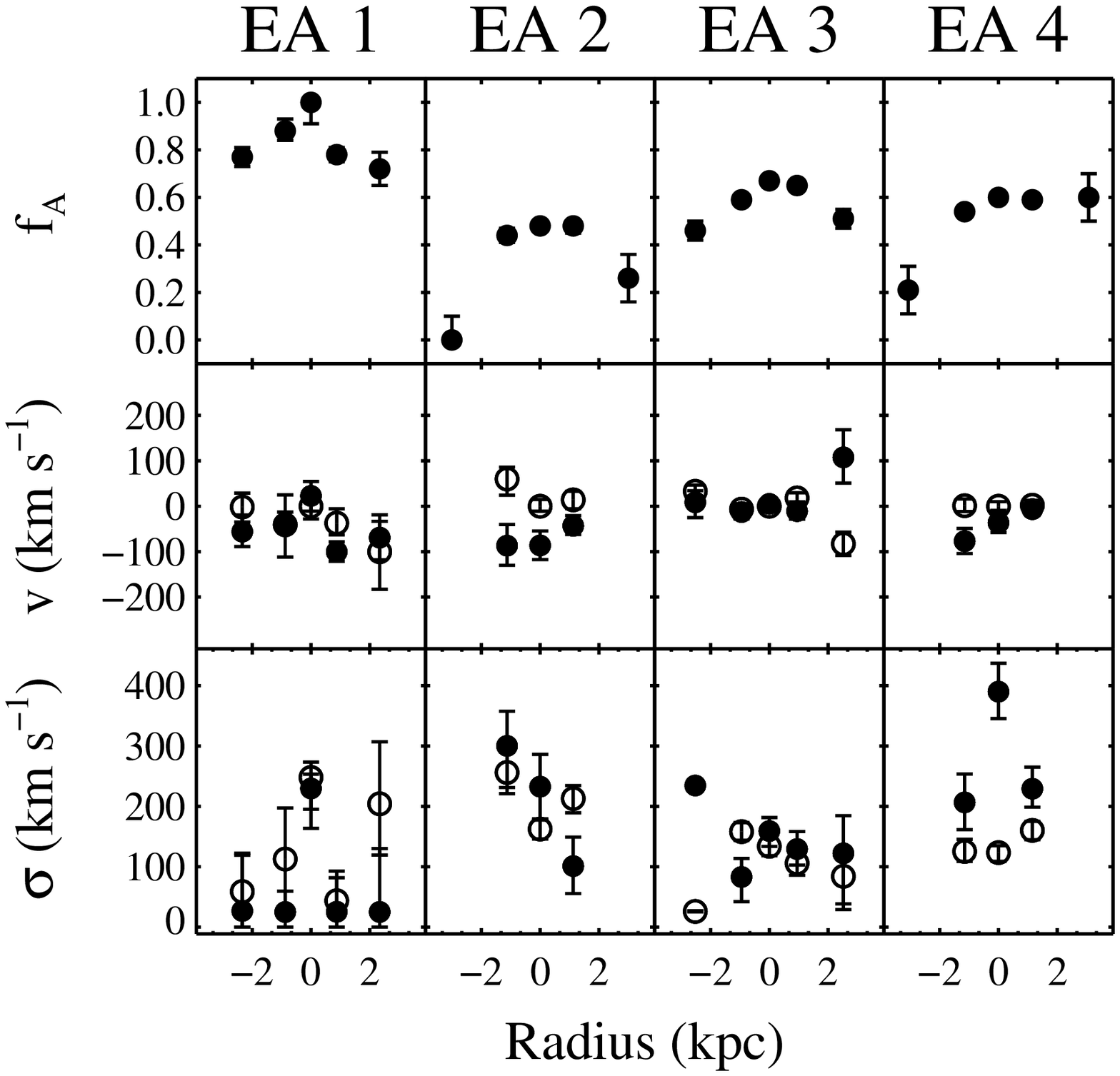}
\caption{Fitted quantities for EA~1 to 4.  Three measurements are
plotted vs. radius. From upper to lower panels these quantities are
(a) f$_A$, the A-type stellar fraction, (b) the mean velocity, and (c)
the velocity dispersion.  Open symbols represent the old stellar
population, and filled symbols are the young population. The velocity
and velocity dispersion are calculated only for the three central
apertures unless the signal-to-noise in the outer apertures exceeds
10; however, we retain the measurements of A star fraction in the
outer apertures (see text).\label{fig:firstall}}
\end{figure}

\begin{figure}
\plotone{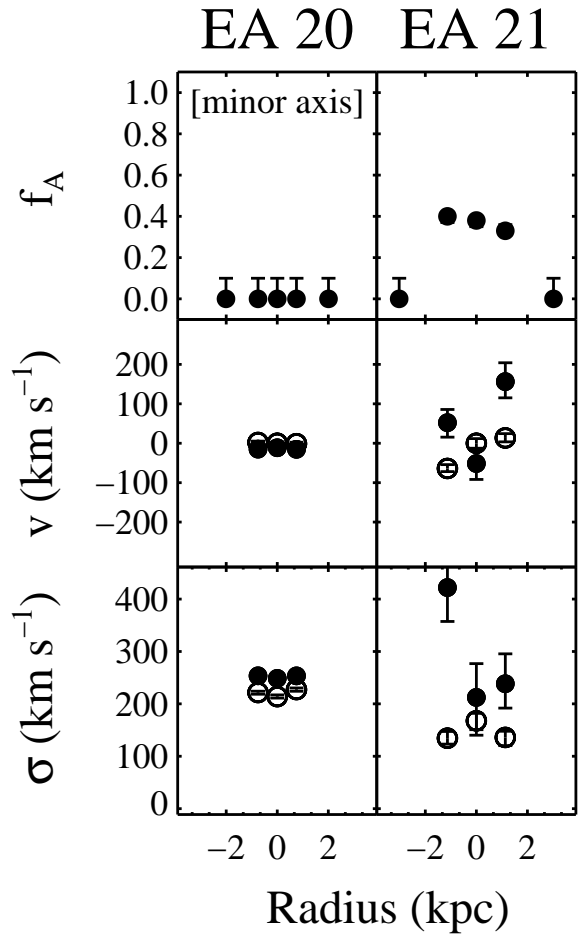}
\caption{Same as Figure~\ref{fig:firstall}, but for the minor axis of EA~20
and for EA~21.\label{fig:lastall}}
\end{figure}

For each aperture along the spectrum of each E+A, we measure the
fraction of light contributed by the young, A-type stars (denoted by
f$_A$), the mean projected velocity relative to the galaxy center
($\langle {\rm v}_{rot} \rangle$), and the velocity dispersion
($\sigma$) (see Figures~\ref{fig:firstall}--\ref{fig:lastall}).
Before examining the properties of the entire sample, we note three
special cases.  First, EA~20 shows no evidence for a young population
in either a major or minor axis spectrum.  We conclude that it was
misclassified as an E+A due to noise in the optical LCRS fiber
spectrum.  We will exclude this galaxy from the rest of the
discussion.  Second, EA~12 has detectable emission in both [OIII] and
Balmer lines.\footnote{This galaxy is also one of two E+As in this
sample detected in radio continuum by Miller et al. (2001, in prep.).}
We exclude the central $\sim$ 9 \AA\ from each Balmer line in fitting
the LOSVD for this galaxy.  Despite the detected emission, the total
[OIII] emission line equivalent width is $< 1$ \AA\ and the average
absorption line equivalent width of H$\beta$, $\gamma$, and $\delta$
is $\sim$ 5.5 \AA.  These parameters indicate that EA~12 does meet the
selection criteria (cf. Z96), and we thus retain this galaxy in our
analysis.  Finally, HST imaging of EA~1 shows two distinct components
that are separated spatially by $\sim$ 2.3 kpc (Zabludoff~\etal\ in
preparation).  We find that the differences between the kinematics of
the two subcomponents are insignificant.  In
Figures~\ref{fig:firstall}--\ref{fig:lastall} and in all future
discussion we refer to the average of the two components as EA~1.  In
Table~\ref{tab:params} we list the value of f$_A$ for the aperture
that includes the entire spectrum, the rotation velocity (see \S
3.2.1), and velocity dispersion for each galaxy.

\begin{figure}
\plotone{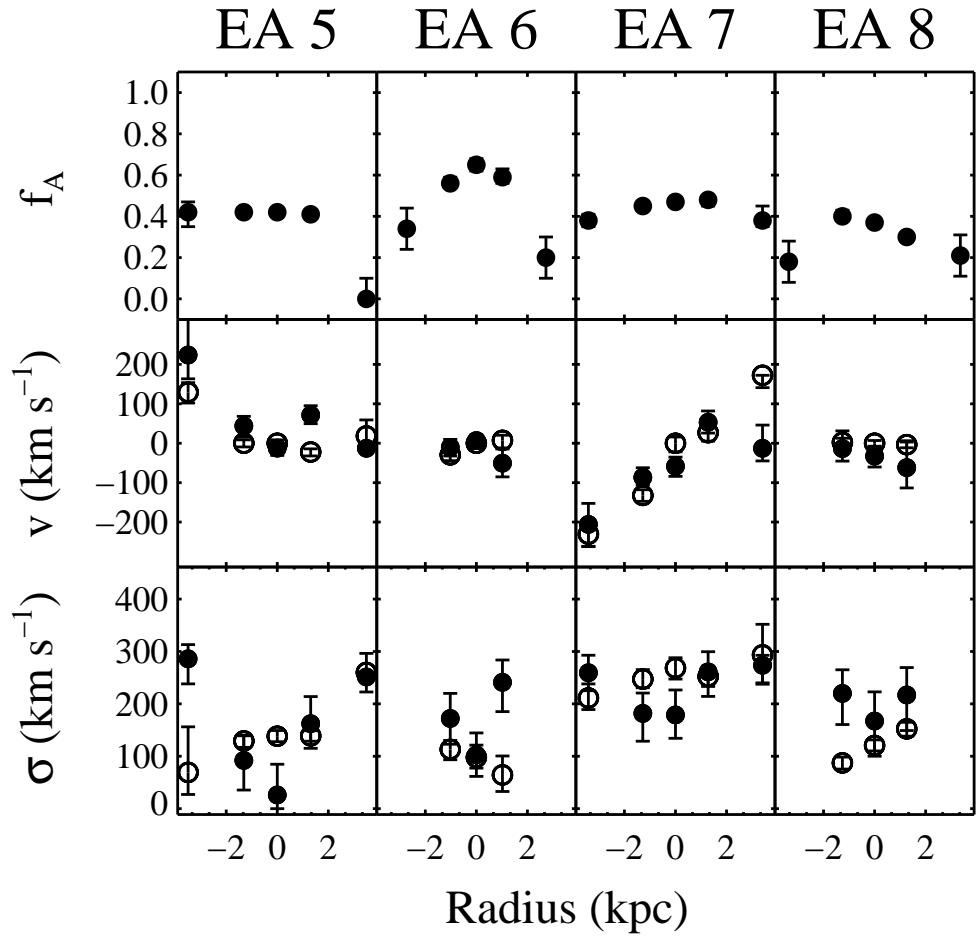}
\caption{Same as Figure~\ref{fig:firstall}, but for EA~5 to 8.}
\end{figure}

\begin{figure}
\plotone{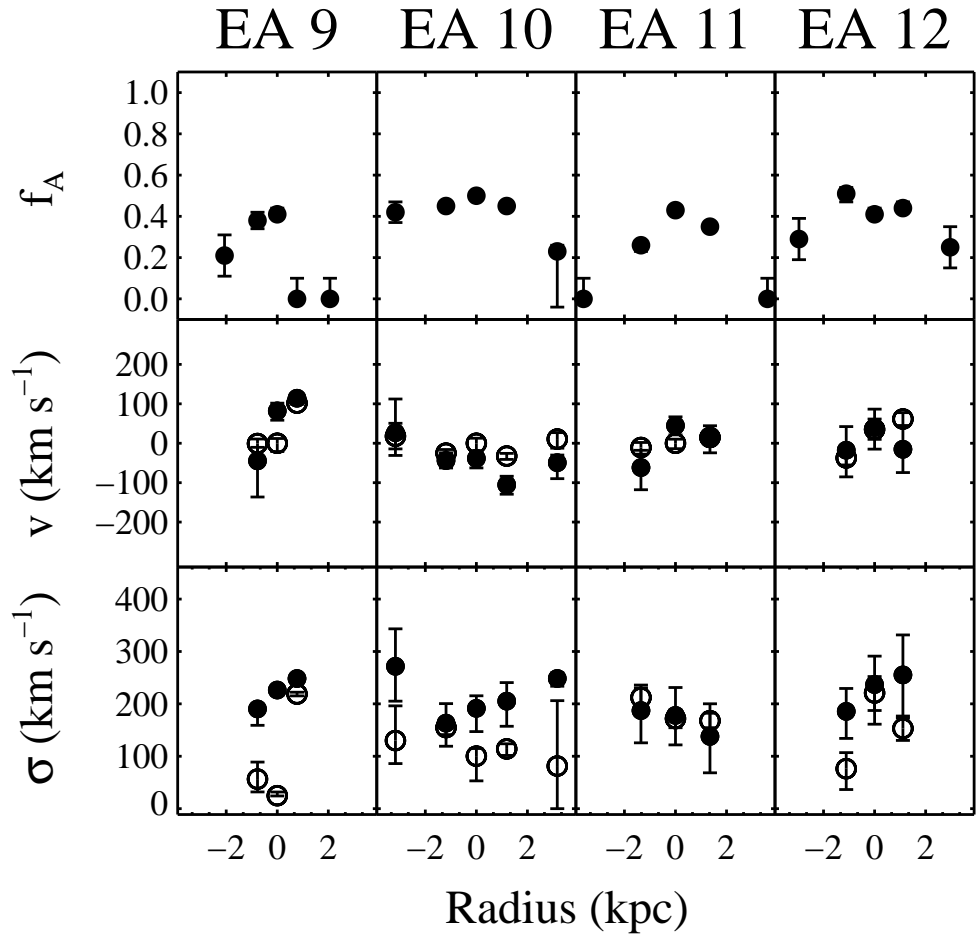}
\caption{Same as Figure~\ref{fig:firstall}, but for EA~9 to 12}
\end{figure}

\begin{figure}
\plotone{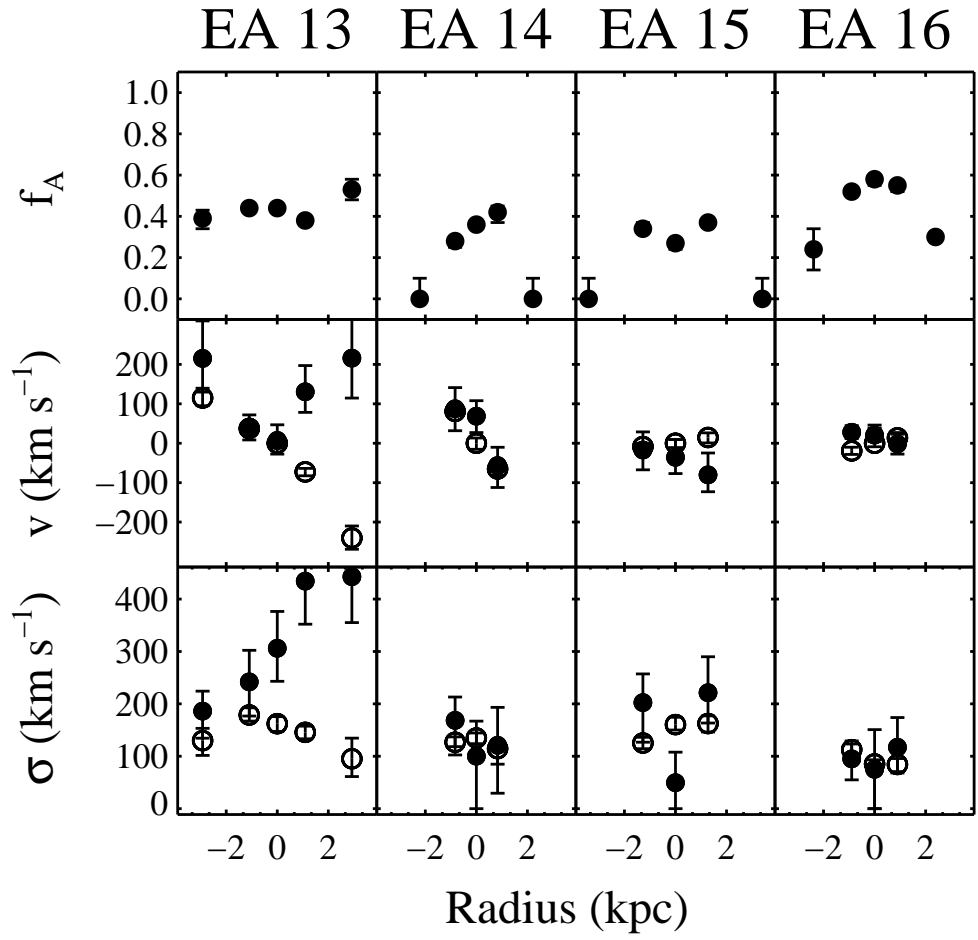}
\caption{Same as Figure~\ref{fig:firstall}, but for EA~13 to 16.}
\end{figure}

\begin{figure}
\plotone{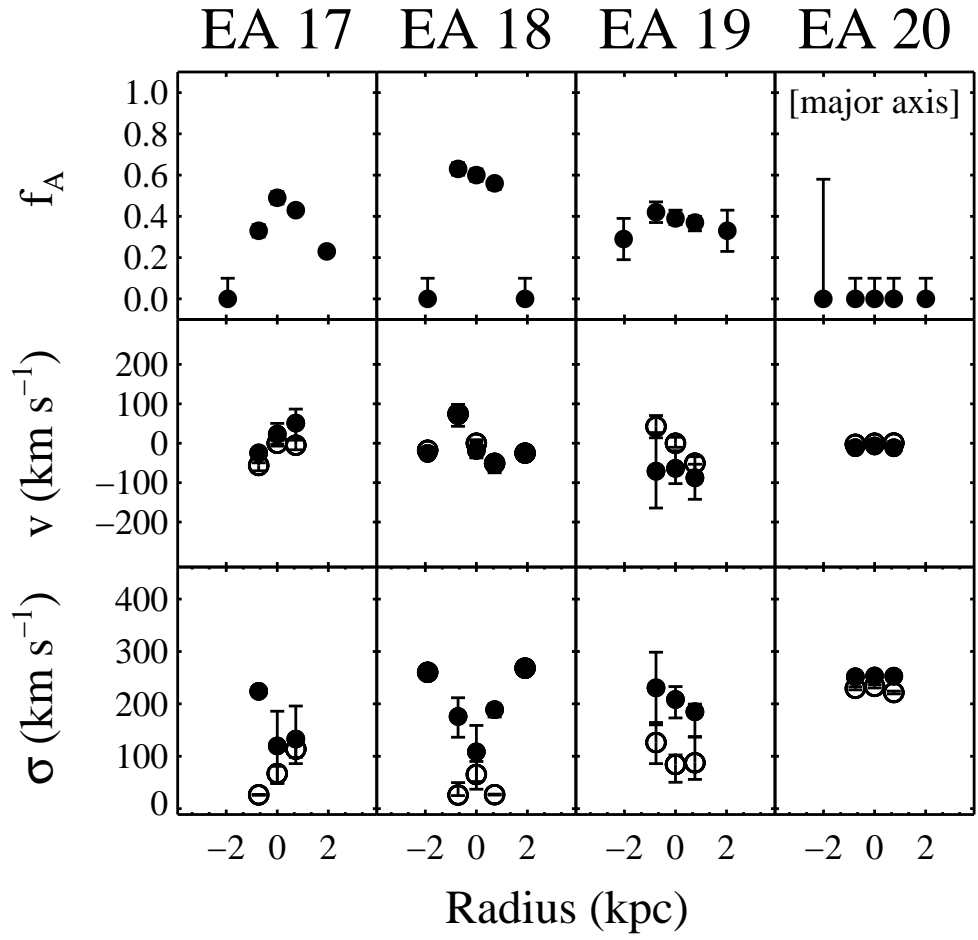}
\caption{Same as Figure~\ref{fig:firstall}, but for EA~17 to 20.}
\end{figure}

\begin{deluxetable}{lccccccc}

\tablecaption{Galaxy Properties\label{tab:params}}

\tablewidth{393.143pt}
\tablehead{
\colhead {Name} & 
\colhead {M$_{\rm R}\tablenotemark{a}$} & 
\colhead {z} & 
\colhead {f$_A$} & 
\multicolumn{2}{c}{$\langle {\rm v}_{rot} \rangle$ (km/s$^{-1}$)} & 
\multicolumn{2}{c}{$\sigma$ (km s$^{-1}$)\tablenotemark{b}} \\
\colhead{} & \colhead{} & \colhead{} & \colhead{} &
\colhead{Old} & \colhead{Young} & 
\colhead{Old} & \colhead{Young} }

\startdata
EA 1 & $-$19.45 & 0.08 & 0.80 $^{+ 0.03}_{- 0.03}$ & 
24 $^{+ 57}_{- 54}$ & 19 $^{+ 17}_{- 17}$ & 
35 $^{+ 39}_{- 35}$ & 23 $^{+ 53}_{- 23}$ \\
EA 2 & $-$20.72 & 0.10 & 0.42 $^{+ 0.02}_{- 0.03}$ & 
23 $^{+ 16}_{- 20}$ & 22 $^{+ 24}_{- 25}$ & 
202 $^{+ 17}_{- 16}$ & 193 $^{+ 32}_{- 46}$ \\
EA 3 & $-$21.73 & 0.08 & 0.59 $^{+ 0.02}_{- 0.03}$ & 
23 $^{+ 17}_{- 23}$ & 25 $^{+ 18}_{- 17}$ & 
120 $^{+ 22}_{- 20}$ & 56 $^{+ 35}_{- 32}$ \\
EA 4 & $-$21.45 & 0.11 & 0.56 $^{+ 0.01}_{- 0.01}$ & 
1 $^{+ 8}_{- 9}$ & 35 $^{+ 15}_{- 16}$ & 
131 $^{+ 9}_{- 9}$ & 246 $^{+ 24}_{- 25}$ \\
EA 5 & $-$20.73 & 0.13 & 0.42 $^{+ 0.02}_{- 0.01}$ & 
33 $^{+ 24}_{- 25}$ & 52 $^{+ 34}_{- 18}$ & 
120 $^{+ 7}_{- 8}$ & 94 $^{+ 34}_{- 41}$ \\
EA 6 & $-$20.06 & 0.09 & 0.61 $^{+ 0.02}_{- 0.03}$ & 
18 $^{+ 10}_{- 13}$ & 20 $^{+ 20}_{- 22}$ & 
23 $^{+ 1}_{- 23}$ & 158 $^{+ 14}_{- 18}$ \\
EA 7 & $-$22.40 & 0.12 & 0.46 $^{+ 0.01}_{- 0.01}$ & 
141 $^{+ 20}_{- 22}$ & 83 $^{+ 22}_{- 18}$ & 
241 $^{+ 11}_{- 12}$ & 208 $^{+ 30}_{- 31}$ \\
EA 8 & $-$20.45 & 0.12 & 0.36 $^{+ 0.02}_{- 0.01}$ & 
3 $^{+ 7}_{- 7}$ & 24 $^{+ 34}_{- 36}$ & 
99 $^{+ 8}_{- 11}$ & 141 $^{+ 55}_{- 55}$ \\
EA 9 & $-$18.74 & 0.07 & 0.27 $^{+ 0.05}_{- 0.26}$ & 
51 $^{+ 19}_{- 6}$ & 79 $^{+ 46}_{- 17}$ & 
74 $^{+ 134}_{- 5}$ & 172 $^{+ 4}_{- 60}$ \\
EA 10 & $-$20.53 & 0.11 & 0.46 $^{+ 0.01}_{- 0.02}$ & 
4 $^{+ 21}_{- 19}$ & 34 $^{+ 25}_{- 17}$ & 
121 $^{+ 7}_{- 8}$ & 183 $^{+ 34}_{- 41}$ \\
EA 11 & $-$20.64 & 0.13 & 0.33 $^{+ 0.01}_{- 0.03}$ & 
13 $^{+ 10}_{- 8}$ & 38 $^{+ 32}_{- 29}$ & 
174 $^{+ 12}_{- 11}$ & 95 $^{+ 52}_{- 46}$ \\
EA 12 & $-$20.30 & 0.10 & 0.50 $^{+ 0.03}_{- 0.03}$ & 
49 $^{+ 12}_{- 11}$ & 1 $^{+ 45}_{- 42}$ & 
111 $^{+ 17}_{- 18}$ & 470 $^{+ 69}_{- 78}$ \\
EA 13 & $-$21.20 & 0.10 & 0.40 $^{+ 0.02}_{- 0.01}$ & 
116 $^{+ 20}_{- 20}$ & 22 $^{+ 42}_{- 38}$ & 
165 $^{+ 10}_{- 9}$ & 392 $^{+ 66}_{- 59}$ \\
EA 14 & $-$20.40 & 0.07 & 0.35 $^{+ 0.02}_{- 0.02}$ & 
73 $^{+ 9}_{- 9}$ & 72 $^{+ 39}_{- 36}$ & 
135 $^{+ 10}_{- 10}$ & 73 $^{+ 50}_{- 73}$ \\
EA 15 & $-$20.30 & 0.12 & 0.33 $^{+ 0.02}_{- 0.02}$ & 
12 $^{+ 8}_{- 8}$ & 32 $^{+ 31}_{- 37}$ & 
125 $^{+ 9}_{- 10}$ & 187 $^{+ 56}_{- 57}$ \\
EA 16 & $-$19.87 & 0.08 & 0.51 $^{+ 0.01}_{- 0.01}$ & 
16 $^{+ 8}_{- 7}$ & 15 $^{+ 15}_{- 15}$ & 
93 $^{+ 7}_{- 8}$ & 95 $^{+ 36}_{- 29}$ \\
EA 17 & $-$19.05 & 0.06 & 0.51 $^{+ 0.05}_{- 0.04}$ & 
26 $^{+ 9}_{- 7}$ & 38 $^{+ 18}_{- 21}$ & 
23 $^{+ 1}_{- 23}$ & 189 $^{+ 9}_{- 8}$ \\
EA 18 & $-$19.90 & 0.06 & 0.61 $^{+ 0.02}_{- 0.02}$ & 
33 $^{+ 14}_{- 9}$ & 31 $^{+ 8}_{- 8}$ & 
23 $^{+ 4}_{- 23}$ & 168 $^{+ 16}_{- 20}$ \\
EA 19 & $-$19.70 & 0.07 & 0.39 $^{+ 0.04}_{- 0.03}$ & 
46 $^{+ 17}_{- 16}$ & 8 $^{+ 56}_{- 50}$ & 
84 $^{+ 18}_{- 34}$ & 208 $^{+ 25}_{- 35}$ \\
EA 20 & $-$20.15 & 0.07 & 0.00 $^{+ 0.10}_{- 0.00}$ & 
2 $^{+ 2}_{- 2}$ & 
\nodata & 204 $^{+ 2}_{- 2}$ & \nodata \\ 
EA 20\tablenotemark{c} & $-$20.15 & 0.07 & 0.00 $^{+ 0.10}_{- 0.00}$ & 
2 $^{+ 2}_{- 2}$ & 
\nodata & 200 $^{+ 2}_{- 2}$ & \nodata \\ 
EA 21 & $-$20.22 & 0.10 & 0.35 $^{+ 0.02}_{- 0.02}$ & 
39 $^{+ 7}_{- 7}$ & 52 $^{+ 30}_{- 27}$ & 
134 $^{+ 7}_{- 9}$ & 237 $^{+ 48}_{- 45}$ \\
\enddata

\tablenotetext{a}{Magnitudes are from the LCRS survey \citep{linetal96}; errors are $\sim$ 0.1 magnitude.}
\tablenotetext{b}{$\sigma$ = 23 km s$^{-1}$ is the limit of our resolution; all entries with $\sigma$ = 23 km s$^{-1}$ are upper limits.}
\tablenotetext{c}{Slit aligned with minor axis.}
\end{deluxetable}


\subsection{Stellar population gradients}
\label{sec:populations}

The variation of f$_A$ across the slit measures the distribution of
the recent star formation episode, which in turn constrains star
formation and merger models (cf. \citet{js92}; \citet{mh96}).  A trend
is apparent in the behavior of f$_A$ along the slit
(Figures~\ref{fig:firstall}--\ref{fig:lastall}, top panels): the young
star fraction typically peaks toward the center of the galaxy.  Of the
exceptions to this trend, EA~5, EA~7, EA~13, and EA~19 have a nearly
constant f$_A$, and EA~12 and EA~15 have a slight decline in f$_A$ at
the nucleus. For EA~12 at least, the nuclear minimum in f$_A$ may be
due to contamination by Balmer line emission, which is particularly
prominent near the galaxy center.  We have attempted to exclude this
emission from our fits as described above, but some effects could
remain.

\begin{figure}
\epsscale{.85}
\plotone{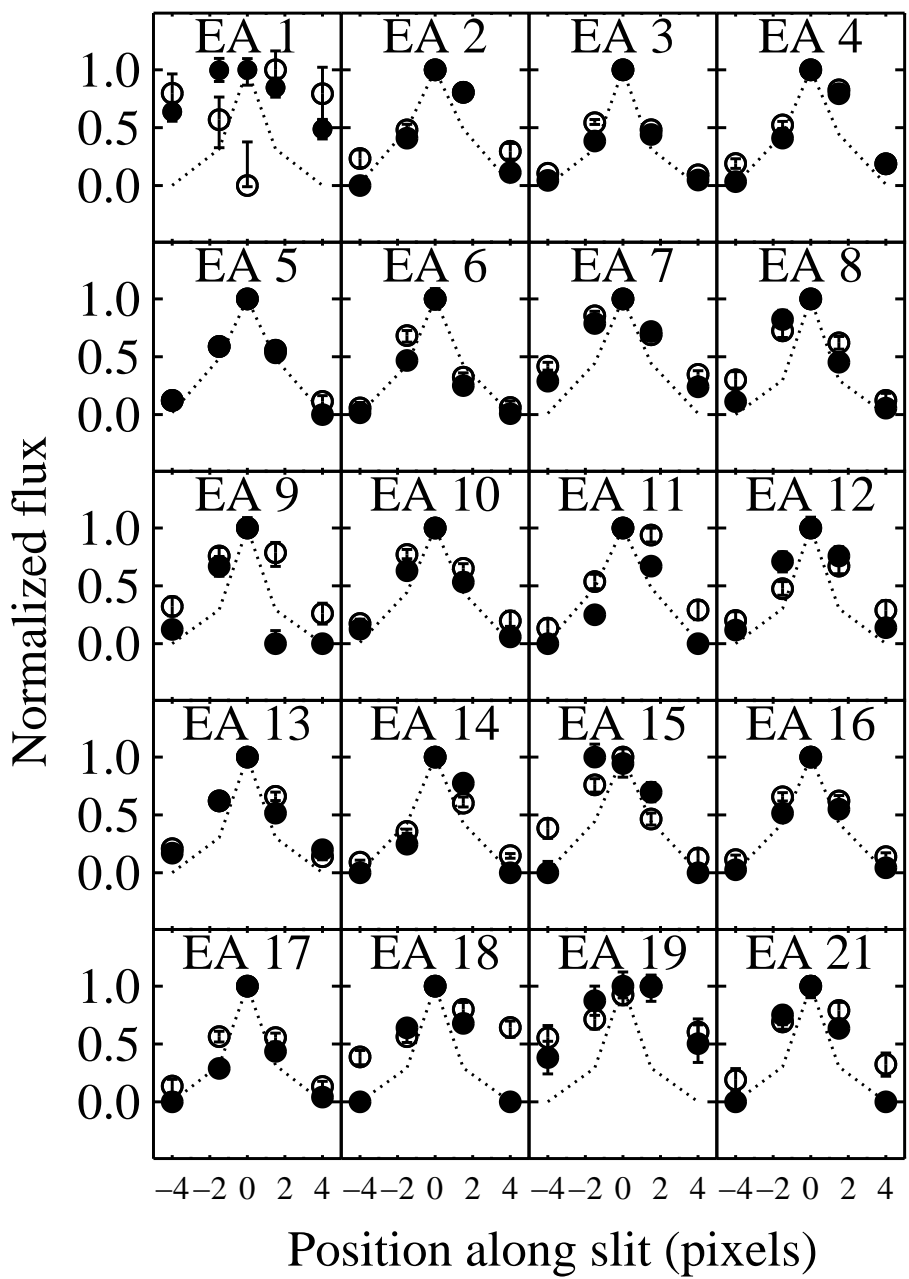}
\caption{Flux contributed by each population as a
function of radius, normalized at the central pixel.  The filled circles
represent the young stellar population, and the open circles are
the older population. The dotted line represents a Gaussian seeing
profile measured from a corresponding stellar image (see text).
\label{fig:akflux}}
\end{figure}

The second moment of the luminosity distributions measures the
relative extent of the populations.  Because our
observations span angular distances of only $\sim$ 2$\arcsec$ to 
7$\arcsec$ (or $\sim$
2 to 7 kpc) across the galaxy, a range which is roughly comparable to the
seeing (typically 1.5$\arcsec$), a serious concern
is that the light of the young stars that we observe outside the nuclear
region is simply scattered nuclear light. To address this concern,
we compare the distribution
of light from the stellar populations of each galaxy with the seeing
profile in Figure~\ref{fig:akflux}. 
The seeing profile is obtained from stellar spectra that were taken
within a couple of hours before and after each galaxy spectrum.  We
overplot a Gaussian profile of the corresponding width (which ranges
from 1.2$\arcsec$ to 1.7$\arcsec$ FWHM) for each galaxy in
Figure~\ref{fig:akflux}.

We find that 6 of the 20 E+A galaxies have flux from young stars that
is more than 2$\sigma$ above the Gaussian profile at {\it both} of the
outermost measured points (EA~1, 3, 7, 12, 13, and 19).  
For comparison, the identical analysis for
the old stellar population flux shows that 15 of the galaxies display
a 2$\sigma$ flux excess relative to a nuclear source at the outermost radii.
The remaining galaxies, EA~5, 6, 10, 15, and 21, also show old stellar
populations that are more extended than the Gaussian, but not at the
2$\sigma$ level.

\begin{figure}
\plotone{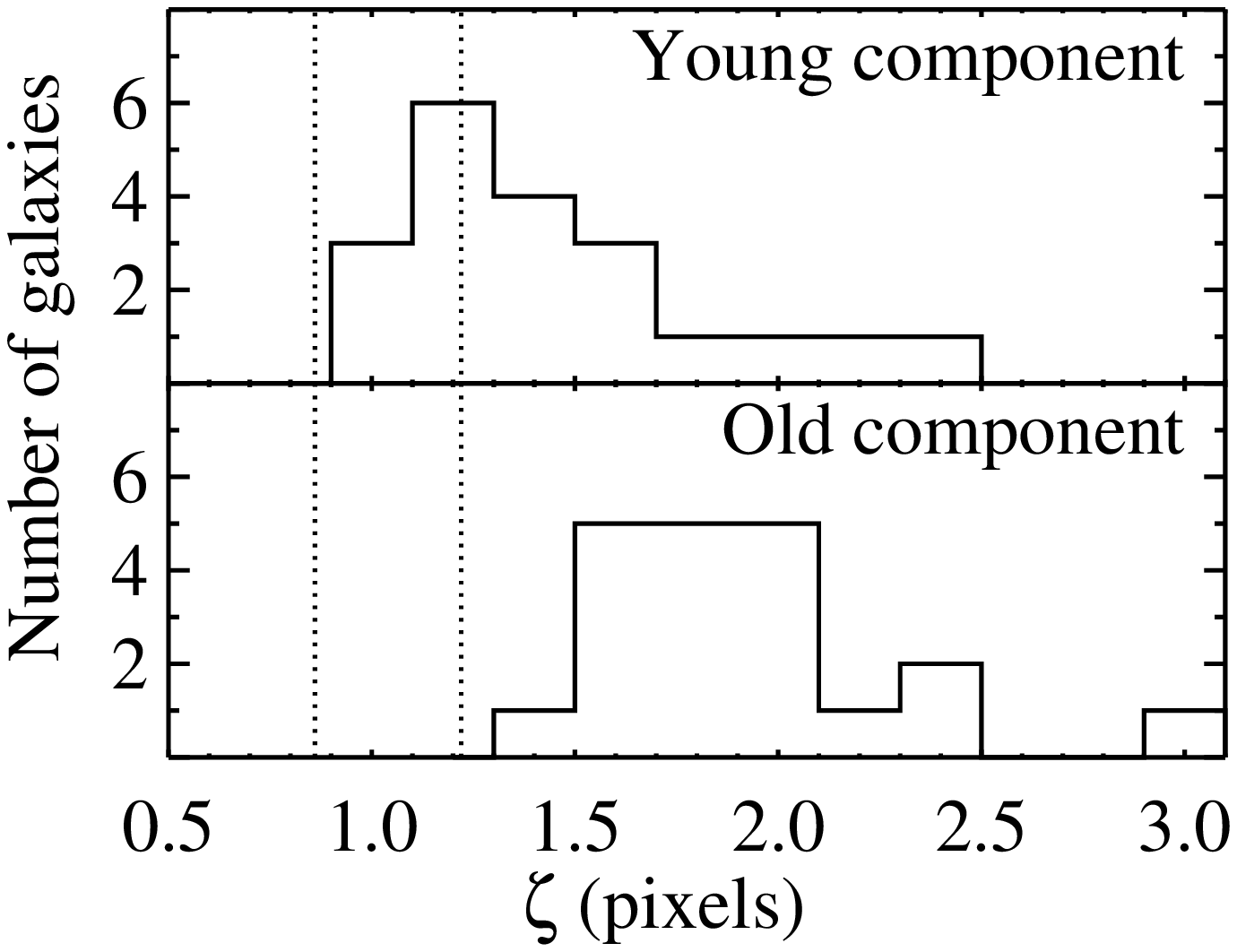}
\caption{Histogram of the second moment of the luminosity
distribution ($\zeta$) for the young and old stellar populations.  The
dotted lines mark $\zeta$ for a Gaussian point spread function with
the maximum and minimum seeing widths observed during the run.
\label{fig:afrachist}}
\end{figure}

To quantify these results further, in Figure~\ref{fig:afrachist} we
plot the second moment, $\zeta$, of each galaxy's luminosity
distribution. For a normal distribution, $\zeta$ is the FWHM of the
luminosity distribution divided by 2.35.  Two vertical lines mark the
range of $\zeta$ determined from the seeing profiles over the course
of our observations. About one half of the young stellar populations
are marginally resolved while the old populations are well resolved.
Seven E+As (EA~1, 3, 7, 8, 12, 13, and 19) have resolved, extended
young populations, with $\zeta$ exceeding the seeing by more than
40\%.  Only one of the 20 E+A galaxies, EA~12, has a detectable young
stellar population that is (marginally) more extended then its older
population, but its central emission may have some effect on this result.

From our study of the stellar population distributions, we conclude
that 
(1) the older stellar populations in these galaxies are
more radially extended than the young stellar populations
and (2) the recent star formation in E+A galaxies, though it is
relatively centrally concentrated, is not always confined to the
galaxy's central kiloparsec.


\subsection{Velocity Distributions}

Stellar velocity distributions are one of the fundamental differences
between different galaxy types. In this Section we present and 
discuss our measurements of the first and second moments of the
velocity distributions of E+A galaxies.

\subsubsection{Mean Velocities}

We begin by examining the first velocity moment (mean velocity) as a
function of position along the slit. Monotonic variations along the
slit would indicate rotation.  It must be noted that these data are
only able to probe rotation at distances of $\sim$ 1 -- 3 kpc, near
the galaxy centers.

From Figures~\ref{fig:firstall}--\ref{fig:lastall} it is evident that
there is generally no {\em strong} rotation, in either the old or
young populations of these galaxies.  The most notable exception is
EA~7, in which both stellar populations rotate at $\sim$ 200 km
s$^{-1}$ at a radius of $\sim$ 4 kpc. There is a suggestion of
rotation in a few others, such as the old stellar populations of EA~13
and 14 and the young populations of EA~9 and 14. However, aside from
these few exceptions, we find little evidence for significant rotation
in the E+As.

We define the characteristic rotation velocity $\langle {\rm v}_{rot}
\rangle$ of a galaxy to be equal to one half the velocity difference
across the slit. We average the velocities in the two radial bins on
each side of the slit for higher signal-to-noise.  The resulting
rotation velocity is not directly comparable to the full rotation at
the rotation curve turnover radius.  For a solid-body rotation curve
over the radial range probed, $\langle {\rm v}_{rot} \rangle$ would be the
velocity at the average distance of our intermediate radial bins,
$\sim$ 1.9 kpc.  For reference, typical spiral galaxies have $\langle
{\rm v}_{rot} \rangle \sim (70 {\rm\ to\ } 140 {\rm \ km\ s^{-1}})/\sin{i}$
(when measured at the same radius; see \eg,\ \citet{hsmp}), where $i$
is the unknown inclination angle.

Only six of the twenty E+As in our sample have $\langle {\rm v}_{rot}
\rangle > 50$\ km s$^{-1}$ for either the old or the young stellar
populations (Table 2).  Using a Kolmogorov-Smirnov (K-S) test, we find
that the distributions of $\langle {\rm v}_{rot} \rangle$ of the young and
old stellar populations are not significantly different. Not only are
the distributions similar, but, for most of the E+As, rotation is
observed either in both populations or not at all.  For example, only
four E+As have old stellar populations with $\langle {\rm v}_{rot} \rangle >
50$ km s$^{-1}$: EA~7, EA~9, EA~13, and EA~14.  Of these four
galaxies, three (EA~7, 9, and 14) also show evidence for rotation in
the young stellar component. Only two galaxies (EA~7 and 13) have
differences larger than 50 km s$^{-1}$ in $\langle {\rm v}_{rot} \rangle$
between the young and old populations. However, interpreting the
results from such comparisons, whether they are similarities or
differences, is complicated by the different spatial distribution of
the two populations.

An alternative explanation for the monotonic change in velocity across
the slit is bulk relative motion of two merging galaxies.  However,
none of the rotating E+As show separated optical components on the
Digitized Sky Survey images (see Z96), and only one of the five E+As
for which we have HST imaging (EA~1; Zabludoff~\etal\ in preparation)
has two components (and EA~1 shows no apparent rotation).

The majority of our sample does not appear to be rapidly rotating.
Two factors may lead us to underestimate the degree of rotation in our
sample.  First, our rotation velocities must be corrected by sin $i$
to account for inclination effects.  However, it is unlikely that many
of the galaxies would be fortuitously aligned close to face-on.
Assuming a random distribution of inclinations, such that the
probability of observing a galaxy with inclination $i < \theta$ is
$1-{\rm cos}(\theta)$, a random distribution of galaxy velocities
between 40 and 100 km s$^{-1}$ --- a very conservative estimate ---
would result in a $\langle {\rm v}_{rot} \rangle$ distribution that has
$\sim$ 25\% of the measurements below 40 km s$^{-1}$.  In our sample,
75\% of the galaxies have velocities below this threshold, suggesting
that the rotation velocities of the E+As are smaller than the 40 to
100 km s$^{-1}$ range.  Second, poor spatial resolution may be a
concern.  If E+A galaxies were rotationally supported only outside the
central $\sim$ 2 kpc, we would not be able to observe that here.  It
may be significant that we do detect rotation in some of the galaxies
for which our radial coverage is greatest (EA~3, EA~7). Although we
cannot rule out rotation at larger radii than we are currently able to
probe, $\langle {\rm v}_{rot} \rangle \sin i$, even at the small radii we
observe, is 70 to 140 km s$^{-1}$ for a typical disk galaxy
\citep{hsmp} --- much larger than what we observe in the E+A sample.

From the rotation analysis, we conclude that (1) both rapidly rotating
systems (cf. \citet{franx93} and discussion above) and slowly or
non-rotating systems can be found in the E+A phase, (2) most E+A
galaxies in our sample (14/20) show no evidence for rotation and that
for two other galaxies the evidence for rotation is marginal, and (3)
the rotation (or lack thereof) is generally present in both the
old and young stellar populations of an E+A galaxy.


\subsubsection{Velocity Dispersions}

The lack of rotation in most of the E+As in our sample indirectly
suggests that most are pressure-supported.  To assess this
possibility, we measure the velocity dispersion, $\sigma$, and compare the
dispersions of the young and old stellar populations. Differences
between the two populations may arise if they have different spatial
distributions, if one or both stellar populations are not yet in
equilibrium after the hypothesized merger event, or if they have
reached different dynamical equilibria.

\begin{figure}
\plotone{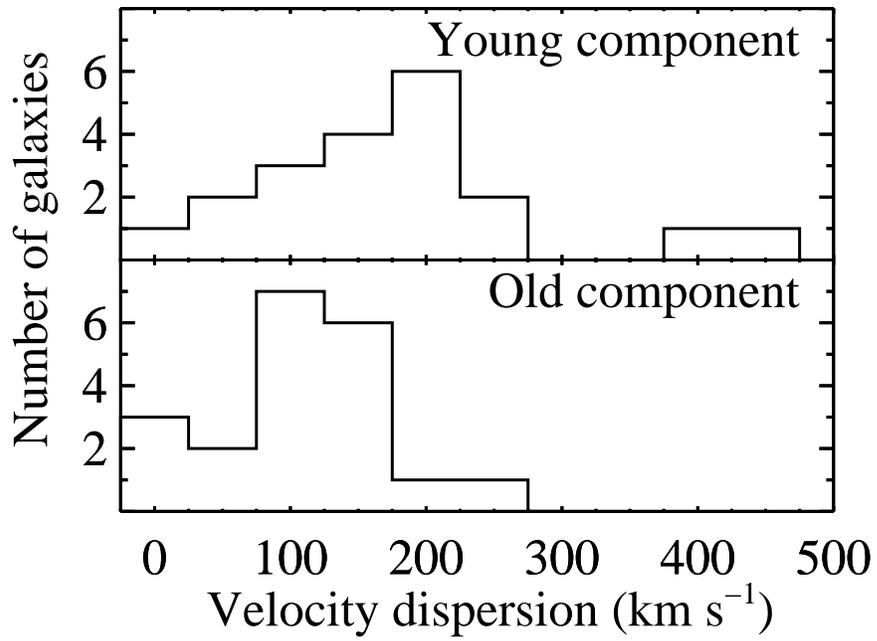}
\caption{Histogram of velocity dispersion in the ``total galaxy''
aperture for each stellar population.  On average, the younger
population has a larger dispersion (see text for details and
caveats). \label{fig:sighist}}
\end{figure}

In Figure~\ref{fig:sighist}, we present a histogram of $\sigma$
calculated from the ``total galaxy'' apertures of $\sim$ 7 kpc in full
width. Because this measure includes line-broadening due to rotation,
the use of a ``total'' aperture is justified only for the low
rotational velocities measured in the majority of these systems
($\sigma$ will be somewhat inflated in systems with significant
rotation, $\langle {\rm v}_{rot} \rangle >$ 50 km s$^{-1}$).  A K-S test
indicates that the $\sigma$'s for the young and old populations are
unlikely ($<$ 1\% chance) to come from the same parent
distribution. To conservatively quantify the difference between the
two populations, we first exclude the four lowest-luminosity E+As, for
which the old populations' velocity dispersions are extremely low.  We
then calculate that the mean values of $\sigma$, and the errors in the
means, are 190 $\pm 12$ km s$^{-1}$ for the young population, and 130
$\pm 5$ km s$^{-1}$ for the old population.  Although the difference
between $\sigma$ in the two populations is significant and argues that
the younger component is more pressure-supported, it may be due to the
aforementioned spatial differences between the populations.  Because
the young stars are more centrally concentrated, and $\sigma$ for a
typical elliptical is larger toward the center (\eg,\
\citet{davies81}), the larger velocity dispersions for the young stars
may be a result of their position within the galaxies.  In normal
ellipticals, varying the aperture from the central kiloparsec outward
can decrease the observed velocity dispersion up to 25\%, although the
typical decline is 5 to 10\% \citep{davies81,getal00}.  From these
data, we are unable to determine whether the observed velocity
dispersion differences are caused by spatial or dynamical differences.

\begin{figure}
\epsscale{0.8}
\plotone{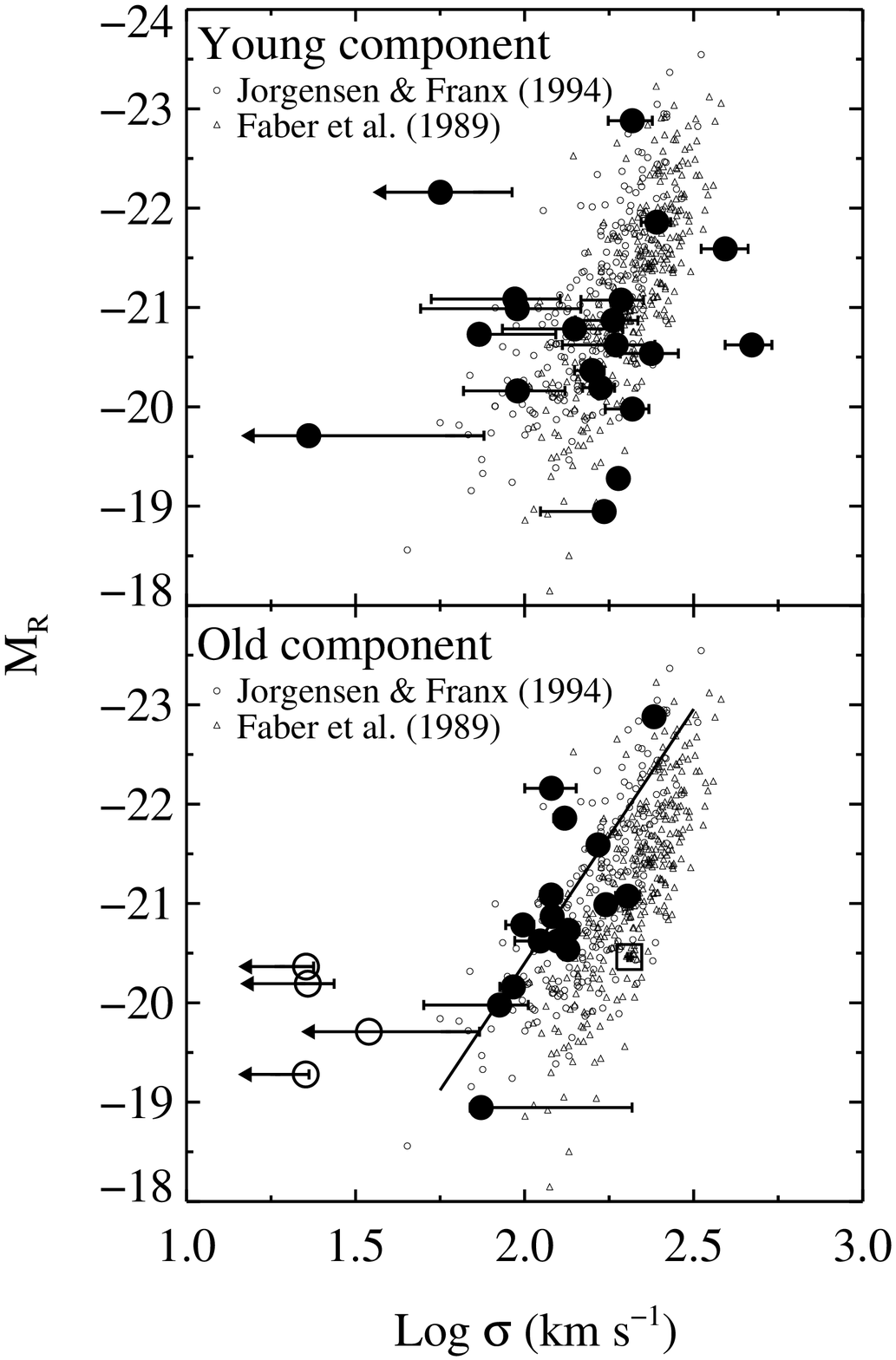}
\caption{The luminosity-velocity dispersion relation for the E+A
galaxies (open and filled large circles). The solid line represents a
least-squares fit, weighting all points equally, to the filled points
in the bottom panel; four outliers (open circles) are excluded from
the fit. The large open square denotes EA~20, which has no young
component and is not truly an E+A galaxy
(\S~\ref{sec:results}).  Also plotted are normal elliptical galaxies
from two studies (small circles and triangles; \citet{fetal89},
\citet{jf94}, respectively). \label{fig:fj}} \epsscale{0.9}
\end{figure}

To compare the properties of the E+As to those of ``normal''
ellipticals, we draw a comparison sample from the studies of
\citet{fetal89} and \citet{jf94}.  The Faber~\etal\ data are published
in the B-band, which we transform to R using a typical B-R color for
this sample of 1.25 (determined from a comparison of galaxies that are
in both the Faber~\etal\ and the R-band J\o rgensen \& Franx data).
In Figure~\ref{fig:fj}, we plot the Faber-Jackson (FJ) relation
\citep{fj76} for these normal ellipticals and compare it to the FJ relations
produced using either the kinematics of the young or old stellar
populations of E+As.  The R magnitudes are for the entire E+A galaxy
and are on a Gunn/Kron-Cousins hybrid system used by the LCRS group
\citep{linetal96}, who report that the differences between the
standard and hybrid Kron-Cousins systems are small ($\lesssim$ 0.1
magnitude). For comparison with the literature data, we have adjusted
the isophotal magnitudes of the E+A galaxies to approximate
total magnitudes by using the systematic trends between the
LCRS isophotal magnitudes and the Sloan Digital Sky Survey 
Petrosian magnitudes, the latter of which
are approximately total magnitudes \citep{bl01}.

There is no significant correlation between galaxy luminosity and
the velocity dispersion of the younger populations in E+As.
Nevertheless, with the exception of a few outliers (EA~12 and 13,
which have extremely high velocity dispersions, and EA~1 and EA~3,
with anomalously low velocity dispersions), the distribution of points
is consistent with the Faber-Jackson relationship.  All other galaxies
are within the envelope formed by the ``normal'' ellipticals, although
the scatter is visibly much larger than that for the ``normal"
ellipticals.

On the other hand, the old stellar populations have a significant
correlation between galaxy luminosity and velocity dispersion
(Spearman rank-order correlation coefficient of 0.6, with a
significance of 99.8\%). We plot the best-fitting linear relation for the E+As
for comparison, weighting all points equally, but excluding the four
galaxies with only upper limits on $\sigma$. The scatter about this
line, 0.6 mag, is the same as that around the
relation for normal ellipticals (we measure a dispersion of 0.64 mag
about the mean relation for the combined comparison sample).
The faintest galaxies (M$_{\rm R} \gtrsim -20$), at least
those with only upper limits on $\sigma$, are
the outliers from this relation. 
The four outliers are drawn from the seven E+As with the largest
values of f$_A$ (including EA~1 and the two other galaxies with the
largest values of f$_A$), suggesting that these galaxies
may harbor more recent, unrelaxed mergers (like EA~1).
On average, excluding these outliers, the E+As lie slightly
to the low-dispersion (or high-luminosity) side of the elliptical galaxy
relation, presumably at least in part because the young stellar populations
elevate the
luminosities. The offset in magnitude from Figure~\ref{fig:fj} is such
that the E+A galaxies are, on average, $\sim$ 0.6 magnitudes brighter
than their elliptical counterparts.  Models \citep{poggianti99,
kelson00} show that over the next few Gyr the young stars in the E+A
galaxies should experience dimming of approximately 0.5 to 1 mag,
after which these galaxies would lie directly on the locus of
elliptical galaxies.

\begin{figure}
\plotone{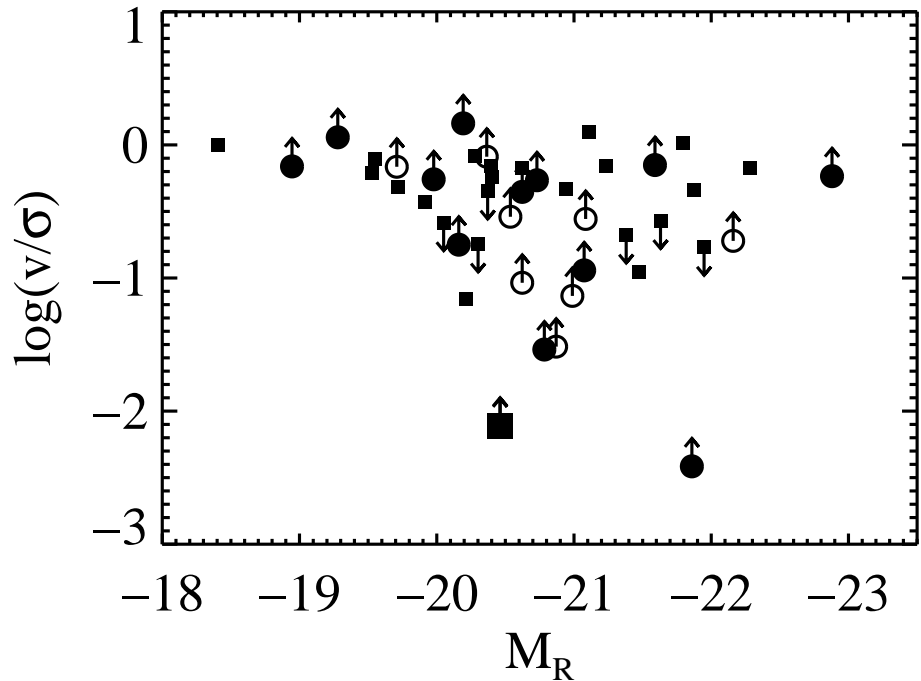}
\caption{v$/\sigma$ plotted against R-band absolute
magnitude.  Large circles (both open and filled) represent our E+A
sample.  Small squares represent elliptical galaxies from \cite{defis}
and \cite{bn90} for which R-band magnitudes are available from the
NASA/IPAC Extragalactic Database (NED). Rotational velocities in our
sample are not corrected for inclination, so our points are marked as
lower limits.  Unfilled circles are galaxies that are only marginally
resolved (see Section~\ref{sec:populations}), but we find no
systematic differences in v$/\sigma$ between these and the other
galaxies in our sample. \label{fig:magvovers}}
\end{figure}

To examine the outliers in relation to the rest of the E+A sample, we
plot the ratio of rotational velocity to velocity dispersion
(v/$\sigma$) versus magnitude (Figure~\ref{fig:magvovers}), adopting
$\langle {\rm v}_{rot} \rangle$ of the old population for v.  As described
in \S3.2.1, the rotation velocities are uncorrected for inclination,
so all v/$\sigma$ points in Figure~\ref{fig:magvovers} are lower
limits. Because we are concerned that spatial resolution may affect
this measurement, we have divided the sample: the eight least-resolved
galaxies (EA~1, 3, 5, 6, 10, 11, 15, and 21, as discussed in
Section~\ref{sec:populations}) are denoted by open circles, and the
remaining galaxies are represented by filled circles. The
poorly-resolved galaxies span a large range in v/$\sigma$, so we
conclude that low spatial resolution, which might cause us to
underestimate the rotational velocities and overestimate velocity
dispersions, is not grossly biasing our results toward low
v$/\sigma$. E+As fainter than M$_{\rm R} \sim -20$ are limited to high
v/$\sigma$, and are thus rotationally supported.  Previous work
\citep{defis, bn90} has similarly found that for M$_{\rm B} \gtsim
-20$, normal ellipticals are flattened by rotation rather than by
anisotropic velocity dispersions. The remainder of the sample (with
M$_{\rm R} < -20$) have an average v$/\sigma$ of 0.32.

From the analysis of velocity dispersions we reach three conclusions.
First, the young populations have on average a higher velocity
dispersion than the old populations.  Three factors which may
contribute to this result are: (a) the spatial segregation of the two
populations (the young stars are more centrally concentrated and
therefore lie deeper in the potential and have a higher velocity
dispersion), (b) a transient, non-relaxed young component (the large
observed $\sigma$ does not yet reflect an equilibrium value), and
(c) a relaxed young component with kinematics distinct from those of
the older component (as would arise if the A stars trace the formation
of a hotter stellar component, such as a bulge).  We cannot yet
differentiate among these alternatives.

Second, the old component of the E+A galaxies conforms to the
Faber-Jackson relation, although the E+As are somewhat brighter ($\sim
0.6$ mag) than typical ellipticals. This result, combined with our
finding that there is little evidence for rotation in most of the
sample, indicates that the galaxies are in general, but not
exclusively, supported by pressure rather than rotation.

Third, we find that the E+A phase is experienced by a wide variety of
galaxies.  The velocity dispersions of E+As in our sample range from
$\lesssim$ 30 km s$^{-1}$ to over 200 km s$^{-1}$.  Some E+As show
significant rotation, but most lie on the Faber-Jackson relation, and
rotational support becomes increasingly important for fainter E+As ---
just as it does for ellipticals.


\section{Discussion}
\label{sec:discussion}

E+A galaxies may represent a short lived phase in the evolution of
many galaxies. Z96 estimated that $\sim$ 0.2\% of all nearby galaxies
are currently observed in an extreme E+A phase. For an estimated
lifetime of the E+A phase of $\sim$ 1 Gyr, their result suggests that
$\gtrsim$ 3\% of galaxies may have experienced such a phase in a
Hubble time.  This is a relatively small fraction of all galaxies, but
if E+As are associated primarily with elliptical galaxies and the
field elliptical fraction is 10\% today, the fraction of ellipticals
that passed through an E+A phase at some point could be as high as
30\%.\footnote{This result is not in conflict with the observational
constraints on the lack of young populations in early type galaxies
\citep{sb98}.  Even if our estimate of 30\% is correct, we would
predict only one galaxy in the Silva \& Bothun sample to have a
significant 1 to 3 Gyr old population, and they observe no galaxies
with such populations contributing at greater than the 10 to 15\%
level.}  This estimate is subject to many uncertainties, but it
demonstrates the potential importance of the E+A phase in galaxy
evolution.  In addition, the number of E+A-like systems over a Hubble
time may actually be underestimated here, because of (1) the extremely
conservative E+A criteria used by Z96, (2) the short lifetime of the
extreme E+A phase (less than 1 Gyr), and (3) the larger fraction of
E+As in the past \citep {dg83, be95}.  As observers probe the
properties of galaxies at earlier and earlier times, it becomes
increasingly important that we understand local counterparts to the
distant post-starburst galaxy population.  E+As provide us with a rare
opportunity to study galaxies in transition.

Our measurements of $\sigma$, v$/\sigma$, and the distributions of
young and old stars indicate that E+As are not, as a class, disk
galaxies that have had their star formation quenched \citep{nbk90} or
that have experienced a burst followed by a cessation of star formation as a result
of interaction with the intracluster medium \citep{dg83}.  However,
one model does not fit all of our E+As. In particular, a few of our
systems have significant ($> 50$ km s$^{-1}$) rotation velocities. Yet
only two of these systems have v$/\sigma ({\rm sin\ } i) > 1$, and so
only a small fraction of the sample is predominantly rotationally
supported. This low fraction seems to be at odds with the high
fraction of such systems among the published kinematic studies of E+As
\citep{franx93,crfl96}.  However, one of the three fast-rotating E+As
in the Coma sample, NGC~4853, has detectable [OII] emission and
similar galaxies have been excluded from the Z96 sample.  The other
two Coma cluster fast-rotators have M$_{\rm R} > -20$, and so are
comparable to the four low-luminosity, rotationally-supported E+As in
our sample. Because of these circumstances, we conclude that our
results are not in disagreement with previously published studies.

Increasingly, it is generally accepted that at least some fraction of
E+A galaxies are the result of a galaxy merger (\eg,\ \citet{lh88,schweizer96}, Z96). There is an abundance of
circumstantial evidence supporting this scenario, from the
observations of tidal tails in several galaxies in this sample (Z96)
to the kinematic observations we present that suggest that E+As are a
link between star forming galaxies (which are likely gas-rich disks)
and quiescent, spheroidal galaxies that lie on the Faber-Jackson
relation and have low v$/\sigma$.  In a merger model for the formation
of E+As, our observation that the young stars are more concentrated
toward the galaxy center than are the older stars indicates that the
gas that formed the younger stars dissipated energy before the
starburst.  Such dissipation is required for the formation of
elliptical galaxies from merging spirals, in order to explain how the cores of
elliptical galaxies increased their phase space density beyond than of their
spiral progenitors \citep{ks92}.  However, that dissipation does not
drive all of the gas to the galaxy nucleus as seen in some simulations
(cf. \cite{mh96}), because the burst of star formation extends beyond
the nucleus in at least seven of our E+As. The disagreement between
observations and simulations hints at a problem with angular momentum
transport or feedback in the simulations \citep{mh94}.  Alternatives
have been suggested (\eg,\ \citet{mbr93}) in which the star formation
is due to molecular cloud collisions \citep{noguchi88, noguchi91}.
Unfortunately, the results of such models are not yet directly
comparable to our observations.

A complete model for the phenomena associated with this phase of
galaxy evolution would relate the evolution of the morphological and
dynamical properties of the galaxy to the evolution of its stellar
populations.  As such, it may at first appear difficult to effect the
hypothesized dramatic structural changes on the short timescales
implied by the stellar populations ($\ltsim 1$ Gyr).  However, violent
relaxation \citep{lyndenbell67} in mergers may reduce the timescale
over which one expects global morphological changes to the order of 10
crossing times ($\sim 10^8-10^9$ yr; \citet{borne84,mihos95}).
Gaseous dissipation can also significantly shorten this timescale
\citep{mrb92}.  Furthermore, the age constraint from stellar
populations reflects the time since the last burst of star formation,
which may come well after any initial interaction between progenitors
(see \citet{mbr93} and \citet{bekki01} for examples of simulations
that predict multiple starbursts in merging systems).  Recent models
that take these effects into account (e.g., \citet{mihos99}) predict
merger dynamical timescales of $\sim$ 1-2 Gyr.  Therefore, although
the synchronization of the dynamical and stellar timescales will
continue to be a challenge for detailed modeling, we conclude that the
two are fully reconcilable.

\section{Conclusions}
\label{sec:conclusions}

We obtain and analyze long-slit spectra to study the spatial
distribution and kinematic properties of the old and young stellar
populations of a local sample of E+A galaxies 
(Z96).  We summarize our results from the study of these
20 E+A galaxies.\footnote{One of the original 21 galaxies, EA~20,
was apparently misclassified due to low S/N in the original LCRS
spectrum.} We have four principal findings:

\begin{enumerate}

\item{The star formation responsible for the young stellar populations
in E+A galaxies is more centrally
concentrated than the older stellar populations in most E+A galaxies in
our sample, but is {\em resolved} beyond the nuclear region in 
seven cases.}

\item{The E+A galaxies in our sample are principally
pressure-supported.  We find that only six of our 20 E+As are rotating
with $\langle {\rm v}_{rot} \rangle \sin i\gtrsim$ 50 km s$^{-1}$ in either the young or
old populations.  Though we are only able to measure the velocities at
radii of 1 -- 3 kpc, this is significantly less rotation than is
typically observed at the same radius in normal spiral galaxies.  We
also find that the old stellar populations of E+A galaxies conform
well to the Faber-Jackson relation. This result argues against the
model in which E+As, as a class, are normal spiral galaxies in which
star formation has recently been extinguished.  On average, the E+As
are slightly ($\sim$ 0.6 mag) brighter than a sample of typical
elliptical galaxies for the same velocity dispersions.  As the young
populations in the E+As age and fade, this luminosity difference will
disappear.  The degree to which the E+As are {\em already} consistent
with the Faber-Jackson relation demonstrates that E+A galaxies relax
quickly ($\ltsim$ 1 Gyr) to the locus of points occupied by typical
ellipticals, which is consistent with the predictions of merger
models.}

\item{We find that the E+A phase of galaxy evolution does not only
produce bright spheroids, but galaxies that span the full range of the
Faber-Jackson relation.  Our sample includes a population of E+As that
resembles the low-luminosity class of ellipticals that are flattened
by rotation \citep{defis, bn90} and produced in high mass-ratio merger
simulations \citep{cr01}.  Given the number of E+As and the duration
of the E+A phase, a simple estimate shows that one-third of field
ellipticals could have passed through this phase at some point.  This
estimate, though it is not expected to predict the actual number of
ellipticals that were once E+As, demonstrates the potential
importance of this phase.  Because E+As populate the entire
Faber-Jackson relation, we conclude that the E+A phase may be one
through which many spheroid-dominated galaxies pass.}

\item{Our kinematic observations are strong evidence that luminous
(M$_{\rm R} < -20$) E+A galaxies are an evolutionary link between
gas-rich, star-forming galaxies and gas-poor, quiescent,
pressure-supported galaxies.  Other authors have highlighted the
spectral evidence for significant recent star formation in E+As (which
demonstrates that these galaxies recently had substantial gas
reservoirs) and against ongoing star formation
\citep{cs87,nbk90,schweizer96,chang}.  This study demonstrates that
faded E+A galaxies will resemble early-type galaxies kinematically, as
well as spectroscopically.  The presence of tidal features in many
(Z96) suggests disk-dominated progenitors.  We conclude that at least
most E+A galaxies in our sample are in the midst of a morphological
transition from late-type (star-forming, gas-rich,
rotationally-supported, and disk-dominated) to early-type (quiescent,
gas poor, pressure-supported, spheroid-dominated).}

\end{enumerate}


\acknowledgments

The authors thank Chris Mihos for his comments on interaction
scenarios, John Mulchaey for his observing expertise, and Greg Bothun
for his helpful comments on the manuscript.  We also thank Luc Simard
for introducing us to the NUKECR algorithm which we used to
effectively remove cosmic rays.  This research has made use of the
NASA/IPAC Extragalactic Database (NED) which is operated by the Jet
Propulsion Laboratory, California Institute of Technology, under
contract with the National Aeronautics and Space Administration.  The
data described herein were taken at Las Campanas observatory, a
facility of Carnegie Observatories. AIZ acknowledges financial support
from NASA grant HF-01087.01-96A. DZ acknowledges financial support
from the David and Lucile Packard Foundation, the Alfred P. Sloan
Foundation, a NASA LTSA grant (NAG 5-3501), and an NSF CAREER grant
(AST-9733111).

\clearpage


\end{document}